\documentclass[12pt]{article}
\usepackage[polish,english]{babel}
\usepackage[latin1]{inputenc}
\usepackage{graphicx}
\usepackage{times}
\usepackage[T1]{fontenc}

\def\br{\begin{eqnarray}}
\def\er{\end{eqnarray}}
\def\be{\begin{equation}}
\def\ee{\end{equation}}
\def\({\left(}
\def\){\right)}

\def\rlx{\relax\leavevmode}
\def\IR{\rlx\hbox{\rm I\kern-.18em R}}

\def\u2{\mid u\mid^2}

\begin{document}
\begin{titlepage}

\begin{center}
{\bf Some properties of (3+1) dimensional vortex solutions\\ in the extended $CP^N$ Skyrme-Faddeev model}
\end{center}

\vspace{.5cm}

\begin{center}
{L. A. Ferreira~$^{\star}$, P. Klimas~$^{\star}$ and W. J. Zakrzewski~$^{\dagger}$}

\vspace{.5 in}
\small

\par \vskip .2in \noindent
$^{(\star)}$Instituto de F\'\i sica de S\~ao Carlos; IFSC/USP;\\
Universidade de S\~ao Paulo  \\ 
Caixa Postal 369, CEP 13560-970, S\~ao Carlos-SP, Brazil\\

\par \vskip .2in \noindent
$^{(\dagger)}$~Department of Mathematical Sciences,\\
 University of Durham, Durham DH1 3LE, U.K.

\end{center}

\begin{abstract}
We look at properties of vortex solutions of the extended $CP^N$ Skyrme-Faddeev model.
We show that only holomorphic solutions of the $CP^N$ model are also solutions of the
Skyrme-Faddeev model. As the total energy of these solutions is infinite these
solutions should be interpreted as describing time dependent vortices. We describe their dynamics and, in partcular, point out that one of the terms in the energy density
is related to the Noether charge of the model.

\end{abstract}

\end{titlepage}

\section{Introduction}

	In spite of many efforts the final description of the strong coupling sector of the Yang-Mills theory still remains a challenge for theoretical physicists. An interesting approach to this problem is based on some non-perturbative methods which involve soliton solutions. The importance of solitons  in particle physics has increased significantly since it became clear that they could play a role as suitable normal modes in the description of strong coupling regime for some physical theories \cite{coleman}. The presence of solitons is an evidence of a high degree of symmetries of the model that can be associated with conserved quantities. The relation between symmetries and conserved quantities is often described in terms of the Noether theorem. However, this is not the case for solitons since the symmetries associated with them  are neither symmetries of the Lagrangian  nor of the  related equations of motion. In fact, at the first sight there are no obvious symmetries  that can
  be associated with such a large number of conserved quantities. For this reason they are called  {\it hidden symmetries}. The study of some two dimensional integrable field theories has shed new light on this issue showing that hidden symmetries are, in fact, symmetries of the zero curvature condition known also as the Lax-Zakharov-Shabat equation \cite{lax}.  The question that rises is if there exists a counterpart of such structures in higher dimensional  field theories and, in particular,  in gauge theories in $(3+1)$ dimensions. An answer to such question was given in \cite{afs}  where a  generalization of some ideas of integrability to higher dimensions was  proposed  using the construction of  flat connections on a loop space. Such an approach  has already led to successes in some field theoretic models and so this approach has been reformulated further 
in \cite{afs-review}. Recently \cite{ flym,flgauge} this approach has been used to construct integral formulations of the classical equations of motion
  of Yang-Mills theories in $(3+1)$ dimensions, Chern-Simons theories i
n $(2+1)$ dimensions, and integrable field theories in $(1+1)$ dimensions. The obtained results show some interesting connections between gauge and integrable field theories. For these reasons this approach seems to be a useful tool for dealing with effective models for strongly coupled Yang-Mills theories such as the Skyrme-Faddeev model \cite{sf}.

Recently, two of us \cite{CPNvortex}, have proposed a version of the $(3+1)$ dimensional Skyrme-Faddeev model which differs from the standard Skyrme-Faddeev model in two aspects. The first difference is the target space of the model. In the case of the model discussed in \cite{CPNvortex} the target space is the complex projective space $CP^N$ which is isomorphic to the symmetric space $SU(N+1)/SU(N)\otimes U(1)$. Due to the arguments presented in \cite{Kondo}, for $N\ge2$ this could be an alternative description of the relevant low-energy degrees of freedom which,  in the paper of Kondo \cite{Kondo}, is refered to as the {\it minimal case}. The other case, for which the target space is the coset space $SU(N+1)/U(1)^N$ which is frequently used in the literature, corresponds to the {\it maximal case}. The version of the Skyrme -Faddeev model and its exact vortex solution in the simplest case of the $CP^1$ target space has been already studied before in \cite{vortexlaf}. The second difference is related to the presence of some additional quartic terms in the Lagrangian. If fact, as  shown by H. Gies in \cite{gies},  such quartic terms appear unavoidably in the Wilsonian effective action of the $SU(2)$  Yang-Mills theory calculated up to one loop
level. This observation suggests a possibility of the appearance of similar quartic terms in the effective action of the physically more relevant case of the $SU(3)$ Yang-Mills theory. When compared with the  $N=1$ case, the Skyrme-Faddeev model on the $CP^N$ target space ($N>1$) allows us to add a further quartic term. For the model with $N>1$ this term provides a new contribution to the total energy density in the integrable sector of the model. This extra contribution drops out when $N=1$. It is  important to gain a better understanding of the main consequences of such quartic terms in the model, and this is one of the motivations of
our study which is described in the present paper. 

In this paper we address these issues and study some exact vortex solutions of the extended Skyrme-Faddeev model for the target space $CP^N$. As shown in \cite{CPNvortex} such theory possesses an integrable sector with an infinite number of local conserved currents.  This integrable sector is defined by some specific conditions on the gradients of the fields and some relations among the coupling constants. We consider a large class of solutions of the sector corresponding to field configurations which are arbitrary functions of the variables $z=x^1+ix^2$ and $y_{+}=x^3+x^0$, with $x^{\mu}$, $\mu=0,1,2,3$, being the Cartesian coordinates of the four dimensional Minkowski space-time. Apparently, there are no finite energy solutions inside this class, but there are physically interesting configurations corresponding to the vortices which are parallel to the $x^3$-axis  with waves traveling along them with the speed of light. In this paper we generalize the results of \cite{CPNvortex} by constructing solutions corresponding to many such vortices distributed symmetrically on the $x^1\,x^2$-plane, and in some cases rotating relative to each other. We study some of their properties; such as their topological charges and the dependence of the energy per unit of length on their separation and on the frequency of their rotation.   It is worth mentioning that our vortices are also solutions of the pure $CP^N$ in $(3+1)$ dimensions and so are related to the configurations considered in \cite{fkz1,fkz2}.

The paper is organized as follows. In the next section we briefly discuss the model, its integrable sector and recall some of its exact solutions and also their relation to the 
solutions of the $CP^N$ model. The third section contains new results obtained from the study of the quartic terms. In the fourth section we present some examples of the exact solutions that illustrate the results of the previous section. We  also present a  numerical example for which the position of the minimum of the energy density depends significantly on the values of the parameters of this solution. The paper finished with some conclusions.
 
\section{The extended Skyrme-Faddeev model}

\subsection{General remarks about the model}

The model considered in  \cite{CPNvortex}, is an extension of the $(3+1)$-dimensional Skyrme-Faddeev model on the target space $CP^N$, obtained by the inclusion of two  terms which are quartic in derivatives of the fields.  The original formulation of the model explored the fact that $CP^N$ is a symmetric space, namely $SU(N+1)/SU(N)\otimes U(1)$, and so the fields where parametrized by the `so-called principal' variable $X\(g\)= g\sigma\(g\)^{-1}$, with $g\in SU(N+1)$, and $\sigma$ being the order two automorphism under which the subgroup  $SU(N)\otimes U(1)$ is invariant.   Therefore, one has $X\(g\,h\)=X\(g\)$, if $h\in SU(N)\otimes U(1)$. For the purpose of this paper  it is enough to express the Lagrangian  in terms of the quantity $P_{\mu}$ defined as $X^{-1}\partial_{\mu}X\equiv \sigma\(g\)\,P_{\mu}\,\sigma\(g\)^{-1}$. In this formulation the Lagrangian of the model takes the form
\br
\mathcal{L}=-\frac{M^2}{2}{\rm Tr}(P_{\mu}^2)+\frac{1}{e^2}{\rm Tr}([P_{\mu},P_\nu])^2+\frac{\beta}{2}\left[{\rm Tr}(P_{\mu}^2)\right]^2+\gamma\left[{\rm Tr}(P_{\mu}P_{\nu})\right]^2,\label{lagrangianSF}
\er
where $P_{\mu}$ can be parametrized by the set of complex scalar fields $u_i(x^0,x^1,x^2,x^3)$ in the following way:
\br
P_{\mu}=\frac{2i}{1+u^{\dagger}\cdot u}\left(
\begin{array}{cc}
0_{N\times N} & \Delta\cdot \partial_{\mu}u \\
\partial_{\mu}u^{\dagger}\cdot \Delta & 0
\end{array}
\right)\qquad{\rm with}\qquad u=\left(
\begin{array}{c}
u_1\\
\vdots \\
u_N
\end{array}
\right).
\er
The symbol $\Delta$ denotes the  hermitian matrix $\Delta_{ij}\equiv\vartheta\,\delta_{ij}-\frac{u_iu_j^*}{1+\vartheta}$ where $\vartheta\equiv\sqrt{1+u^{\dagger}\cdot u}$. Thus in this parametrization of the $N$-dimensional complex projective space $CP^N$ we choose to use $N$ complex scalar fields $u_i$.

The first term in (\ref{lagrangianSF}), namely, $-\frac{M^2}{2}{\rm Tr}(P_{\mu}P^{\mu})$ corresponds to the $CP^N$ Lagrangian which, in the standard formulation, takes the form
\be
\mathcal{L}_{CPN}=M^2(D_{\mu}{\cal Z})^{\dagger}D^{\mu}{\cal Z}, \qquad {\cal Z}^{\dagger}\cdot {\cal Z}=1,\label{lagrangianCPN}
\ee
where the covariant derivative is defined as $D_{\mu}{\Psi}\equiv\partial_{\mu}{\Psi}-({\cal Z}^{\dagger}\cdot
\partial_{\mu}{\cal Z}){\Psi }$. In (\ref{lagrangianCPN}) we have also included a dimensional constant $M^2$ since we are interested in comparing the results of the model based on this term alone to those based on  the full model (\ref{lagrangianSF}). This correspondence is established via the parametrization 
\begin{eqnarray}
{\cal Z}=\frac{(1,u_1,\ldots,u_N)}{\sqrt{1+|u_1|^2+\ldots+|u_N|^2}}
\label{udef}
\end{eqnarray} 
for which $\cal Z$,  by definition, satisfy the condition ${\cal Z}^{\dagger}\cdot {\cal Z}=1$. One can check that the part of the Lagrangian $\mathcal{L}$ proportional to $M^2$ is proportional to $\mathcal{L}_{CPN}$ {\it i.e.} $\mathcal{L}_{M^2}=4\mathcal{L}_{CPN}$.

The last three terms  in (\ref{lagrangianCPN}) are quartic in powers of derivatives. Amongst them the term proportional to $1/e^2$ is the standard Skyrme-Faddeev quartic term whereas those proportional to $\beta$ and $\gamma$ constitute an extension of the model. The extension contains all possible quartic terms that can be considered for this model. 
Note that in the case of the $CP^1$ target space the three quartic terms produce only two types of terms, namely $\(\partial_{\mu}u_1\partial^{\mu}u_1^*\)^2$, and $\(\partial_{\mu}u_1\)^2\,\(\partial_{\nu}u_1^*\)^2$. Thus, in this case, out of the three coupling constants $1/e^2$, $\beta$ and $\gamma$, only two are independent.  This statement is not longer true for $N\ge 2$, where the interplay between the Lorentz indices $\mu$ and $\nu$, and the internal indices $i$ and $j$ produces extra independent terms. Hence, in the case $N\ge 2$ both  $\beta$ and $\gamma$ can play a significant role.

One of the most important results presented in \cite{CPNvortex} was the demonstration that the extended Skyrme-Faddeev model (\ref{lagrangianSF}) possesses an integrable sub-sector (sub-model) with an infinite number of conserved currents. The integrability condition has the form
\be
\partial^{\mu}u_i\partial_{\mu}u_j=0,\qquad {\rm for\,\, any} \,\,i,j=1,2,\ldots,N.\label{constraint}
\ee 
In what follows we shall refer to (\ref{constraint}) as {\it the constraints}. As shown in  \cite{CPNvortex} the equations of motion of (\ref{lagrangianSF})  together with the constraints (\ref{constraint}) imply that for any arbitrary functional $G$ of the fields  $u_i$ and $u_i^*$, but not of their derivatives, there exists a current $J_{\mu}^G$ which is conserved, {\it i.e.} $\partial^{\mu}J_{\mu}^G=0$ (see \cite{CPNvortex} for details). 

In addition, it was also shown in \cite{CPNvortex}  that if one imposes a particular relation among the coupling constants; namely: $\beta e^2+\gamma e^2=2$, the integrable sub-model possesses a wide classes of exact solutions. These solutions are given by arbitrary meromorphic functions of the complex variable $z=x^1+ix^2$ and of the real variable $y_+=x^3+x^0$, with $x^{\mu}$, $\mu=0,1,2,3$, being the Cartesian coordinates of the Minkowski space-time. Indeed, the configurations
\be
u_i=u_i(z,y_+),\qquad u_i^*=u_i^*(\bar z,y_+),\qquad \beta e^2+\gamma e^2=2,\label{constraint2}
\ee
satisfy the Euler -Lagrange equations corresponding to (\ref{lagrangianSF}) as well as the constraints (\ref{constraint}). In fact, there is more to it. The configurations (\ref{constraint2}) satisfy the following  equations
\be
\partial^{\mu}\partial_{\mu}u_i=0, \qquad \partial^{\mu}u_i\partial_{\mu}u_j=0,\qquad \partial^{\mu}[\partial_{\nu}u_i\partial_{\mu}u_j]=0\label{expr}.
\ee

Note that solutions of (\ref{constraint2}) are also solutions of the $CP^N$ model in $(3+1)$ dimensions, {\it i.e.} the model defined by the Lagrangian (\ref{lagrangianCPN}) or the first term in (\ref{lagrangianSF}). That  fact has been explored in \cite{fkz1} to construct vortex solutions for the $CP^N$ model. Therefore, a key point to be stressed is the observation that the $CP^N$ model and the integrable sector of the extended Skyrme-Faddeev model share a wide class of solutions given by the solutions of (\ref{expr}).  The vortex-type solutions presented in \cite{fkz1} are more general that those of \cite{CPNvortex}. Moreover, the vortices can be located at any distance
from each other - and so we can study their dependence on this distance (in the previous case this distance was zero as they were located on `top of each other'). In fact we find that the energy per unit length of the vortex depends in a nontrivial way on the distance between any two or more individual vortices. We would like to add, however, that not all solutions of the $CP^N$ model are simultaneously solutions of the extended Skyrme-Faddeev model. The class of the so-called {\it mixed solutions}, $u_i(z,\bar z,y_+)$ of the $CP^N$ model, see also \cite{wojtekbook}, that we have discussed in \cite{fkz2}, does not satisfy the constraint (\ref{constraint}) and so they are not solutions of the integrable sub-model. In this paper we concentrate our attention on the contribution to the total energy density generated by the quartic terms.  We are especially interested in the dependence of the energy per unit length on the distance between vortices.

\subsection{The Hamiltonian}

It is convenient to split the Hamiltonian into three parts
$$
\mathcal{H}_{c}=8M^2(\mathcal{H}^{(1)}+\mathcal{H}^{(2)})+64(\gamma-\beta)\mathcal{H}^{(3)},
$$
where the subscript $c$ implies that  the Hamiltonian is already restricted to the integrable sector by assuming (\ref{constraint2}). The first two contributions are given by
\be
\mathcal{H}^{(1)}\equiv\frac{\partial_{\bar{z}} u^{\dagger}\cdot \Delta^2
  \cdot\partial_z u}{(1+u^{\dagger} \cdot
  u)^2}, \qquad \mathcal{H}^{(2)}\equiv\frac{\partial_{+} u^{\dagger}\cdot \Delta^2
  \cdot\partial_{+} u}{(1+u^{\dagger} \cdot
  u)^2},
\ee
where $\Delta^2_{ij}=(1+u^{\dagger} \cdot u)\delta_{ij}-u_iu_j^*$.
The quartic contribution can be cast in the form
\be
\mathcal{H}^{(3)}\equiv\frac{1}{(1+u^{\dagger}\cdot u)^4}\sum_{i,j,k,l=1}^N\Delta^2_{ij}\Delta^2_{kl}B_{ik}^*B_{jl},
\label{h3def}
\ee
where $B_{jl}\equiv(\partial_{ z}u_j\partial_{y_+}u_l-\partial_{ z}u_l\partial_{y_+}u_j)$ and $B_{ik}^*\equiv(\partial_{\bar z}u_i^*\partial_{y_+}u^*_k-\partial_{\bar z}u_k^*\partial_{y_+}u^*_i)$.
The reduced Hamiltonian $\mathcal{H}_c$ is positive definite for $\gamma\ge\beta$ since each of its contributions $\mathcal{H}^{(a)}$, $a=1,2,3$ is positive definite.  The proof of the positive definiteness is given in \cite{CPNvortex}. 

The  Lagrangian (\ref{lagrangianSF}) is invariant under the phase transformations $u_i\rightarrow e^{i\alpha_i}u_i$, $i=1,\ldots, N$. The associated Noether currents (for the restricted sub-model) are given by
$$
J_{\mu}^{c(i)}=8M^2\tilde\mathcal{J}_{\mu}^{(i)}+64(\gamma-\beta)\mathcal{J}_{\mu}^{(i)},
$$
where we have split the expression of $J_{\mu}^{c(i)}$ into two parts
\br
\tilde\mathcal{J}_{\mu}^{(i)}\equiv\frac{1}{2i}\frac{1}{(1+u^{\dagger} \cdot
  u)^2}\sum_{j=1}^N\left[u_i^*(\Delta^2)_{ij}\partial_{\mu}u_j-\partial_{\mu}u_j^*(\Delta^2)_{ji}u_i\right]
\er
and
\br
\mathcal{J}_{\mu}^{(i)}\equiv&-&\frac{1}{2i}\frac{1}{(1+u^{\dagger} \cdot
  u)^4}\sum_{j,k,l=1}^N\left[(\Delta^2)_{ij}(\Delta^2)_{kl}u_i^*\partial^{\nu}u_k^*(\partial_{\mu}u_j\partial_{\nu}u_l-\partial_{\nu}u_j\partial_{\mu}u_l)\right.\nonumber\\
  &-&(\Delta^2)_{ji}(\Delta^2)_{lk}u_i\partial^{\nu}u_k(\partial_{\mu}u_j^*\partial_{\nu}u_l^*-\partial_{\nu}u_j^*\partial_{\mu}u_l^*)
  \left. \right].
\er

Next we observe that both contributions $\tilde\mathcal{J}_{\mu}^{(i)}$ and $\mathcal{J}_{\mu}^{(i)}$ are conserved independently as a consequence of the fact that in the integrable sector the fields satisfy the relations (\ref{expr}). Indeed, using (\ref{expr}) one can check that $\partial^{\mu} \tilde\mathcal{J}_{\mu}^{(i)}=0$. Since the Noether theorem implies $\partial^{\mu} \tilde\mathcal{J}_{\mu}^{c(i)}=0$ it follows that  $\partial^{\mu} \mathcal{J}_{\mu}^{(i)}=0$. Another way of obtaining this result is to observe that (\ref{expr}) also defines an integrable submodel of the pure $CP^N$ model, and that  $\tilde\mathcal{J}_{\mu}^{(i)}$ is the Noether current of such a model, associated with the symmetry $u_i\rightarrow e^{i\alpha_i}u_i$, $i=1,\ldots, N$. Therefore,  $\tilde\mathcal{J}_{\mu}^{(i)}$ is conserved, and so is $\mathcal{J}_{\mu}^{(i)}$ by the same argument.

Following $\cite{CPNvortex}$ we shall express the restricted Hamiltonian density $\mathcal{H}_c$ in terms of the time component of above mentioned Noether currents. We find that  
\br
\tilde\mathcal{J}_{0}^{(i)}=\frac{1}{2i}\frac{u^{\dagger}\cdot\partial_{+}u-\partial_{+}u^{\dagger}\cdot u}{(1+u^{\dagger} \cdot
  u)^2}
\er
and
\br
\mathcal{J}_{0}^{(i)}&=&\frac{1}{2i}\frac{1}{(1+u^{\dagger} \cdot
  u)^4}\sum_{j,k,l=1}^N\left[(\Delta^2)_{ij}(\Delta^2)_{kl}u_i^*\partial_{\bar z}u_k^*(\partial_{+}u_j\partial_{z}u_l-\partial_{z}u_j\partial_{+}u_l)\right.\nonumber\\
  &-&(\Delta^2)_{ji}(\Delta^2)_{lk}u_i\partial_{z}u_k(\partial_{+}u_j^*\partial_{\bar z}u_l^*-\partial_{\bar z}u_j^*\partial_{+}u_l^*)
  \left. \right].\label{current0}
\er

Thus the Hamiltonian density can be rewritten as
$$
\mathcal{H}_c=8M^2\partial_z\partial_{\bar z}\ln{(1+u^{\dagger}\cdot u)}+8M^2\sum_{i=1}^Nk_i\tilde\mathcal{J}_{0}^{(i)}+64(\gamma-\beta)\sum_{i=1}^Nk_i\mathcal{J}_{0}^{(i)},
$$
where the first term is purely topological and the last two, related to the Noether currents, involve derivatives {\it w.r.t.} $y_+$. Thus it follows that the energy per unit length can be cast in the form
$$
\mathcal{E}=8\pi M^2Q_{\rm Top}+\sum_{i=1}^Nk_i\left(8\pi M^2\tilde\mathcal{Q}^{(i)}+64\pi(\gamma-\beta)\mathcal{Q}^{(i)}\right),
$$
where the expression in the bracket describes the Noether charges $Q^{(i)}$. The topological charge is given by the integral
\br
Q_{\rm Top}\equiv\frac{1}{\pi}\int_{R^2}dx^1dx^2\mathcal{H}^{(1)}
\label{qtopdef}
\er
and in a similar manner we have introduced $\tilde\mathcal{Q}^{(i)}$ and $\mathcal{Q}^{(i)}$ {\it i.e.}
\br
\tilde\mathcal{Q}^{(i)}\equiv\frac{1}{\pi}\int_{R^2}dx^1dx^2\mathcal{H}^{(2)},\qquad \mathcal{Q}^{(i)}\equiv\frac{1}{\pi}\int_{R^2}dx^1dx^2\mathcal{H}^{(3)}
\er
The reason why we split the charges $Q^{(i)}$ into two parts will be made clear in the next section.

\section{The relation between the Noether charges and the to\-po\-lo\-gi\-cal charge}

In this section we restrict our study to the case $N=2$. The reason for this is twofold. Firstly, this case is more interesting from the physical point of view since the Yang-Mills theory is a gauge theory based on the symmetry group $SU(3)$. Secondly, it is the simplest case for which $\mathcal{H}^{(3)}$ is not identically zero. 

In contradistinction to the parts $\mathcal{H}^{(1)}$ and  $\mathcal{H}^{(2)}$, the contribution to the energy density $\mathcal{H}^{(3)}$ contains terms that have both type of derivatives {\it i.e.  w.r.t.} $z$ and $y_+$. The term $\mathcal{H}^{(1)}$ is a total derivative and therefore  is purely topological whereas $\mathcal{H}^{(2)}$ has nothing to do with topology since it does not even contain any derivatives  {\it w.r.t.} $z$ or $\bar z$. From the point of view 
of topology one can, however, expect that $\mathcal{H}^{(3)}$ may have some interesting properties. Indeed, encouraged by the results of our numerical studies we have  managed to establish a relation between the topological charge $Q_{\rm Top}$ and the contribution $\mathcal{Q}^{(i)}$ to the total Noether charge. The topological charge arises in the contribution to the energy per unit length calculated from $\mathcal{H}^{(3)}$ when only one of the functions $u_1$ or $u_2$ depends on $z$. Without any loss of generality we can choose, for instance, $u_1\equiv u_1(z,y_+)$ and $u_2\equiv u_2(y_+)$. One can then check that in this case the expression $\mathcal{H}^{(3)}$, given in (\ref{h3def}), takes the form 
\br
\mathcal{H}^{(3)}=2\frac{\partial_{\bar z}u_1^*\partial_{z}u_1}{(1+u^{\dagger}\cdot u)^3}\partial_{+}u_2^*\partial_{+}u_2.
\er

Denoting $\alpha\equiv 1+\mid u_2\mid^2$, we observe that $\frac{\partial_{\bar z}u_1^*\partial_{z}u_1}{(1+u^{\dagger}\cdot u)^3}=\frac{\partial_{\bar z}u_1^*\partial_{z}u_1}{(\alpha +\mid u_1\mid^2)^3}=-\frac{1}{2}\frac{d\,}{d\alpha}\left[\frac{1}{\alpha}\frac{\partial_{\bar z}v_1^*\partial_{z}v_1}{(1 +\mid v_1\mid^2)^2}\right]$, where we have denoted $v_1\equiv u_1/\sqrt{\alpha}$. When integrating over $x^1$ and $x^2$  we can use the fact that the integral  $\int_{R^2}dx^1dx^2 \frac{\partial_{\bar z}v_1^*\partial_{z}v_1}{(1 +\mid v_1\mid^2)^2}$ is invariant under rescaling of $v_1$ by a constant, since it is a topological quantity. Therefore, we obtain the result
\br
\int_{R^2}dx^1dx^2\mathcal{H}^{(3)}=\pi Q_{\rm Top}\frac{\partial_{+}u_2^*\partial_{+}u_2}{(1+|u_2|^2)^2}. 
\er
where $Q_{\rm Top}$ is defined in (\ref{qtopdef}). 

Thus we see that the energy per unit length for this special choice of scalar fields $u_1$ and $u_2$ is a product of the topological charge and of a function which depends only on $y_+$.
This implies that, for some special solutions,  the topology can play an important role also for the energy of the interaction generated by the quartic term.  

The zero-components of the Noether currents (\ref{current0}) in this case become
\br
\mathcal{J}_0^{(1)}=0,\qquad \mathcal{J}_0^{(2)}=\frac{\partial_{\bar z}u_1^*\partial_{z}u_1}{(1+u^{\dagger}\cdot u)^3}\frac{1}{i}\left[u_2^*\partial_+u_2-u_2\partial_+u_2^*\right].
\er
The $x^1$ and $x^2$ integration then gives
\br
\int_{R^2}dx^1dx^2\mathcal{J}^{(2)}_0\equiv\pi \mathcal{Q}^{(2)}=\pi Q_{\rm Top}\frac{1}{2i}\frac{u_2^*\partial_+u_2-u_2\partial_+u_2^*}{(1+|u_2|^2)^2},
\er
which is, again, a product of the topological charge and of some function of the variable $y_+$. One can easily check that the only dependence on $y_{+}$ of  $u_2(y_+)$  that guarantees the  proportionality of the two last integrals is $u_2(y_+)=a e^{i \lambda y_+}$ where $a$ and $\lambda$ are real constants. In order to see this one can express the complex scalar field $u_2$ in terms of two real fields $R(y_+)$ and $\Phi(y_+)$ {\it i.e.} $u_2=R(y_+)e^{i{\Phi(y_+)}}$. Then it follows from the proportionality condition that
$$
\frac{\partial_{+}u_2^*\partial_{+}u_2}{(1+|u_2|^2)^2}=\lambda \frac{1}{2i}\frac{u_2^*\partial_+u_2-u_2\partial_+u_2^*}{(1+|u_2|^2)^2}
$$
and so that
$$
R'^2+R^2\Phi'^2=\lambda R^2\Phi'.
$$
This equation is satisfied by $R$ being a constant $R=a$ and $\Phi=\lambda y_+$. Then 
\br
\mathcal{Q}^{(2)}=Q_{\rm Top}\frac{\lambda a^2}{(1+a^2)^2}.\label{relation}
\er

It is easy to observe that this is the only case when the ratio of the two charges is a constant. In consequence, the energy contribution can be expressed in terms of the corresponding Noether charge. If one choses, for instance, $u_2(y_+)$ as a real function $u_2=R(y_+)$ then the contribution to the energy per unit length remains  some function of $y_+$, while still keeping its proportionality to $Q_{\rm Top}$, whereas both Noether charges vanish. 

It is clear that the other possible choice, {\it i.e.} $u_1(y_+)$ and $u_2(z,y_+)$, is equivalent to the one discussed above. The third possibility leading to the proportionality of the energy per unit length  $\int dx^1dx^2\mathcal{H}^{(3)}$ to the topological charge $Q_{\rm Top}$ arises when both scalar fields $u_1$ and $u_2$ possess the same dependence on $z$ (or $\bar z$) {\it i.e.} $u_i(z,y_+)=v(z)g_i(y_+)$. This can be easily seen if, instead of parametrization in terms of $u_1$ and $u_2$, one considers ${\cal Z}$, introduced in (\ref{lagrangianCPN}), parametrized as
\be
{\cal Z}=\frac{(f_1,f_2,f_3)}{\sqrt{|f_1|^2+|f_2|^2+|f_3|^2}}.
\label{zfpar}
\ee
 Comparing this with (\ref{udef}) we note that there are several equally good possibilities of the choice of $u_i$ {\it e.g.}  $u_i=f_i/f_3$ or $u_1=f_1/f_2$, $u_2=f_3/f_2$ {\it e.t.c.} 
This demonstrates that two possibilities 
\br
(u_1,u_2,1)\qquad{\rm and}\qquad (1,\tilde u_2,\tilde u_1)=(1,\frac{u_2}{u_1},\frac{1}{u_1})\nonumber
\er 
are totally equivalent. Moreover, this discussion shows that in this new reparametrisation  we are taken back to the previous case $\tilde u_1(z,y_+)=\frac{1}{v(z)g_1(y_+)}$ and $\tilde u_2(z)=\frac{g_2(y_+)}{g_1(y_+)}$. Hence these two results are equivalent.

\section{Some  examples}

In this section we give some examples of solutions that are based on the solution of the $CP^2$ model discussed in \cite{fkz2} and some generalizations of the multi vortex solutions discussed in \cite{CPNvortex}. We concentrate on the topological case (the quartic term becomes a topological one) but in the final subsection we discuss also a numerical solution which is not of this type. For such a solution the minimum of the energy per unit length depends on the coupling constants.

\subsection{Example 1}

For some arbitrary choices of the set of functions $f_i(z,y_+)$ introduced in (\ref{zfpar}), or equivalently $u_i(z,y_+)$, the energy density per unit length is a very complicated function. Therefore it is quite hard to discover what is the relation between the values of the parameters of the solution and the distances between maxima of the energy density. In order to overcome this problem it is convenient to study some cases with a symmetric distribution of maxima of the energy density. The positions of maxima depend on zeros of the functions $f_i(z,y_+)$. When only one of these functions has zeroes then maxima are located at these zeroes. For this reason we first consider the following case  
\begin{eqnarray}
f_1(z,y_+)&=&z^Q+a_1\,z^N\,e^{ik_1y_+}\nonumber\\ \qquad f_2(z,y_+)&=&a_2\,z+a_3\,e^{ik_2y_+}\nonumber\\ \qquad f_3(z,y_+)&=&a_4,\label{sys4}
\end{eqnarray}
where $N\le Q$. 

The power $Q$ is the maximal degree of polynomials in $z$ which so gives $Q_{\rm Top}=Q$. When $a_2=0$ the energy per unit length $\int dx^1dx^2\mathcal{H}^{(3)}$ has to be proportional to the topological charge. Moreover, in this case $u_2=\frac{a_3}{a_4}e^{i k_2y_+}$. According to the previous section the Noether charge and the topological charge are proportional for $a_2=0$. The energy density contributions thus take the form
\br
\mathcal{H}^{(1)}=\frac{Z_{(Q,N)}}{X_{(Q,N)}^2},\qquad\mathcal{H}^{(2)}=\frac{W_{(Q,N)}}{X_{(Q,N)}^2},\qquad\mathcal{H}^{(3)}=\frac{Y_{(Q,N)}}{X_{(Q,N)}^3},\label{hamiltonians}
\er
where 
\begin{eqnarray}
X_{(Q,N)}&\equiv&a_3^2+a_4^2+a_1^2r^{2N}+r^{2Q}+2a_1r^{N+Q}\cos{[k_1y_+-(Q-N)\varphi]},\nonumber\\
Z_{(Q,N)}&\equiv&(a_3^2+a_4^2)\left[N^2a_1^2r^{2N-2}+Q^2r^{2Q-2}\right.\nonumber\\&+&\left.2NQa_1r^{N+Q-2}\cos{[k_1y_+-(Q-N)\varphi]}\right],\nonumber\\
Y_{(Q,N)}&\equiv&\frac{2k_2^2a_3^2a_4^2}{a_3^2+a_4^2}Z_{(Q,N)},\nonumber\\
W_{(Q,N)}&\equiv&a_3^2a_4^2k_2^2+a_1^2[a_4^2k_2^2+a_3^2(k_1-k_2)^2]r^{2N}+a_3^2k_2^2r^{2Q}\nonumber\\
&-&2a_1a_3k_2(k_1-k_2)r^{N+Q}\cos{[k_1y_+-(Q-N)\varphi]}.\nonumber
\end{eqnarray}
There are two special cases when the formulas simplify a lot; namely:
\begin{enumerate}
\item
$N=0$,
\item
$N=Q$.
\end{enumerate}

\subsubsection{The case $N=0$}

For $N=0$ and $a_1\neq0$, each contribution $\mathcal{H}^{(1)}$, $\mathcal{H}^{(2)}$ and $\mathcal{H}^{(3)}$ have exactly $Q$ symmetrically distributed maxima lying on circles with radii $r^{(1)}$, $r^{(2)}$ and $r^{(3)}$ respectively. The angular position of $k$-th maximum is given by 
\begin{eqnarray}
\varphi_k(y_+)=\frac{1}{Q}\left[k_1y_++[2(k-1)+\theta(a_1)]\pi\right]\label{katy},
\end{eqnarray}
where $\theta{(a_1)}$ is the Heaviside step function and $k=1,2,\ldots,Q$. The radii $r^{(a)}$, $a=1,2,3$, can be calculated exactly. We present here only $r^{(1)}$ and $r^{(3)}$ since $r^{(2)}$ is given by a very complicated expression. They take the form
\begin{eqnarray}
r^{(1)}&=&\left[\frac{|a_1|+\sqrt{Q^2a_1^2+(Q^2-1)(a_3^2+a_4^2)}}{Q+1}\right]^{\frac{1}{Q}},\nonumber\\
r^{(3)}&=&\left[\frac{(Q+2)|a_1|+\sqrt{9Q^2a_1^2+4(Q-1)(2Q+1)(a_3^2+a_4^2)}}{2(2Q+1)}\right]^{\frac{1}{Q}}.\nonumber
\end{eqnarray}

\begin{figure}
\begin{center}
\includegraphics[width=0.3\textwidth]{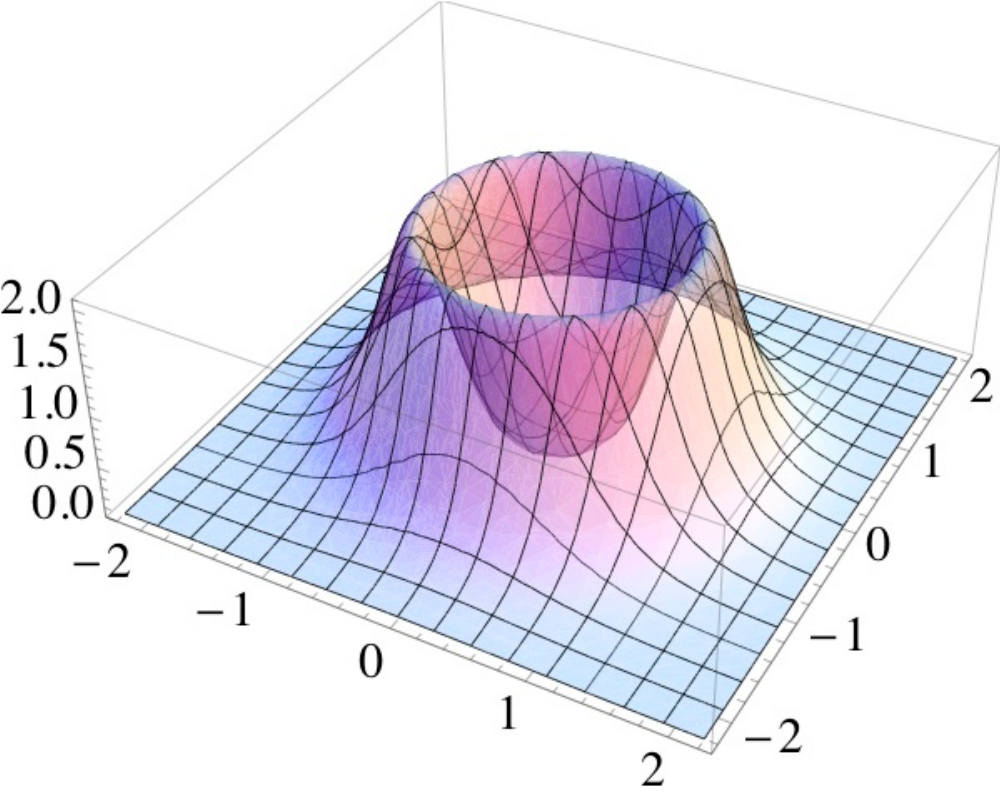}\includegraphics[width=0.3\textwidth]{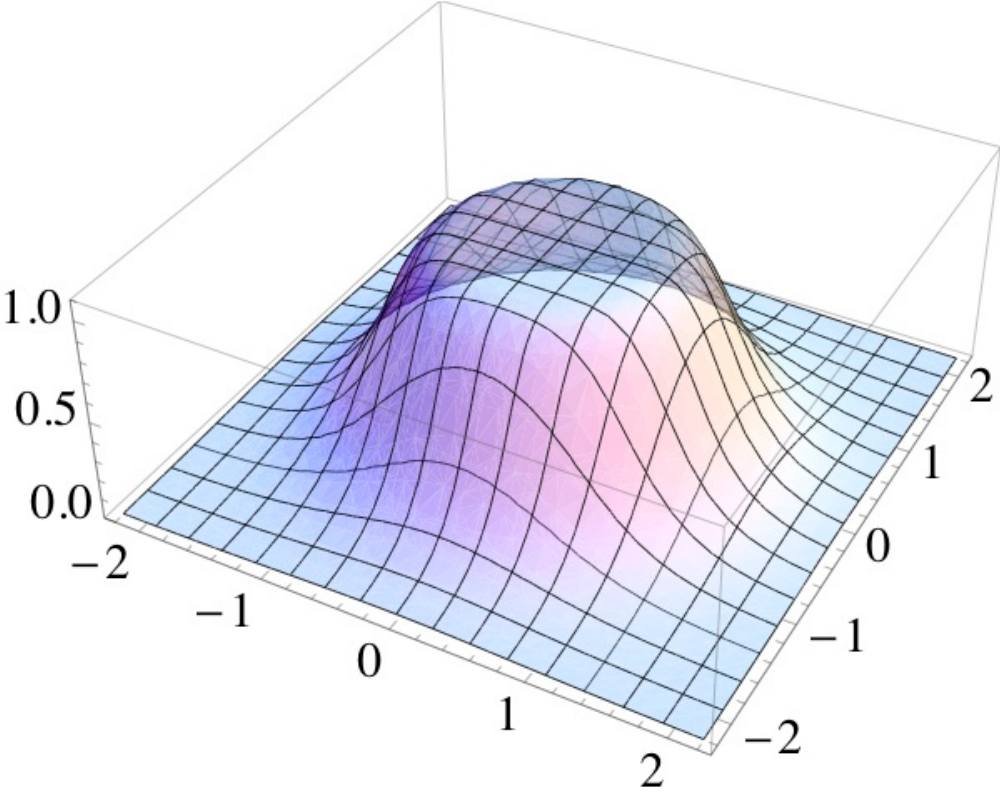}\includegraphics[width=0.3\textwidth]{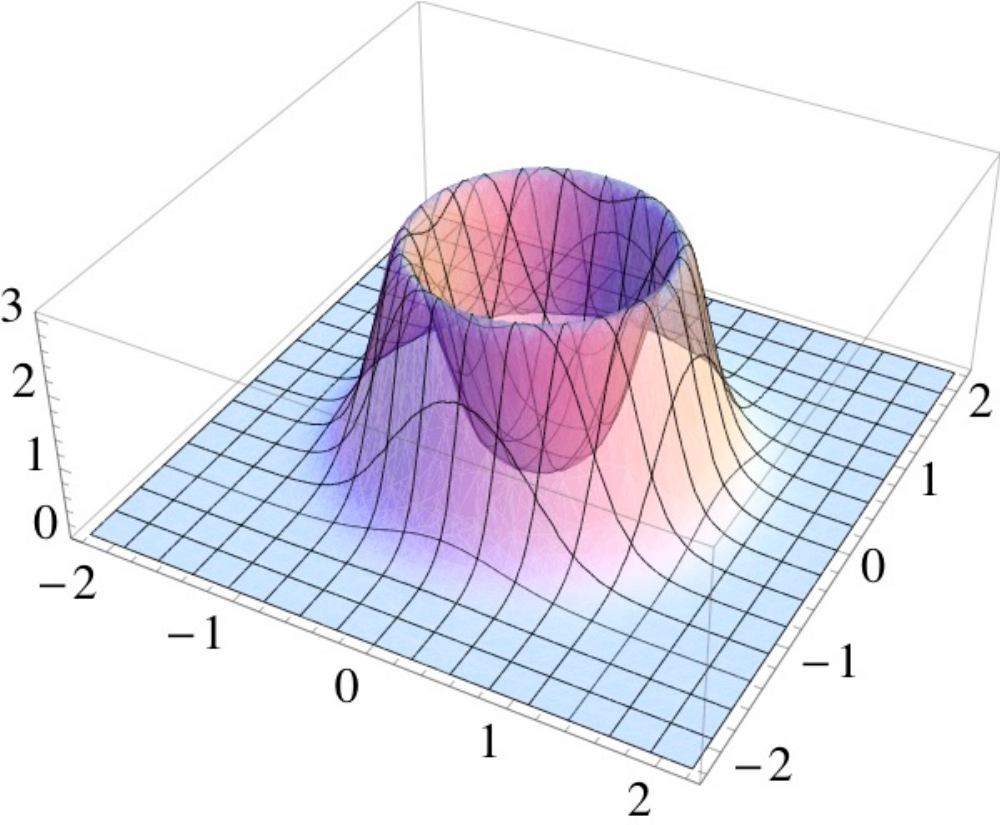}
\includegraphics[width=0.3\textwidth]{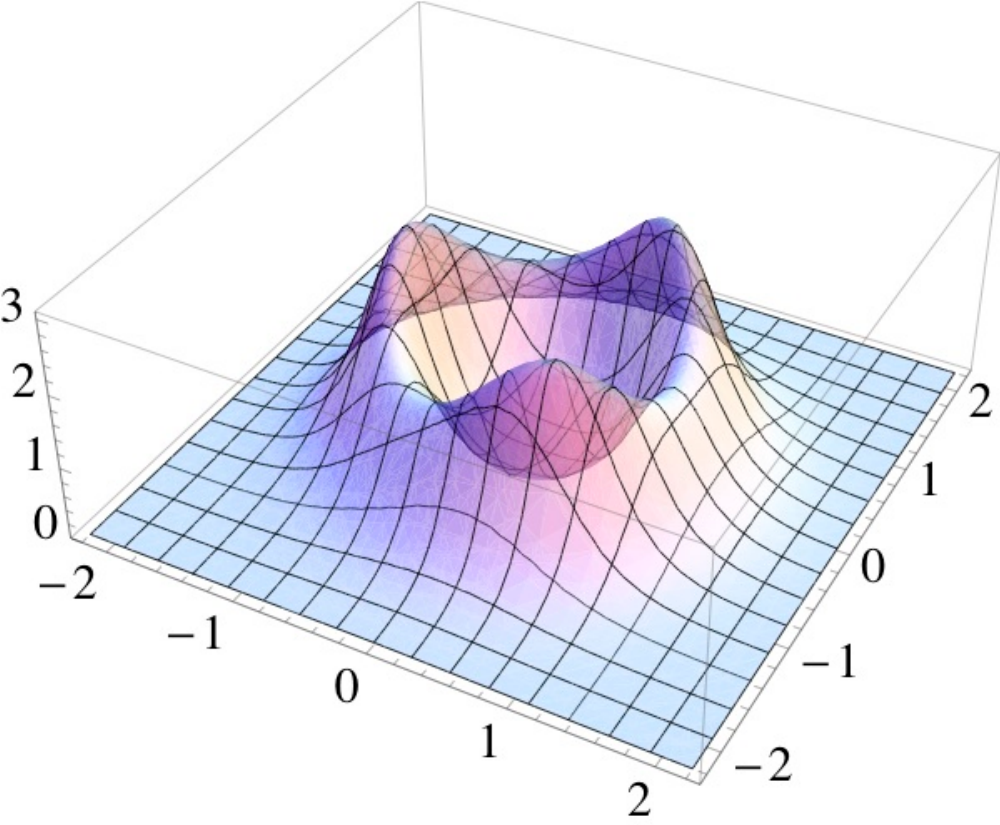}\includegraphics[width=0.3\textwidth]{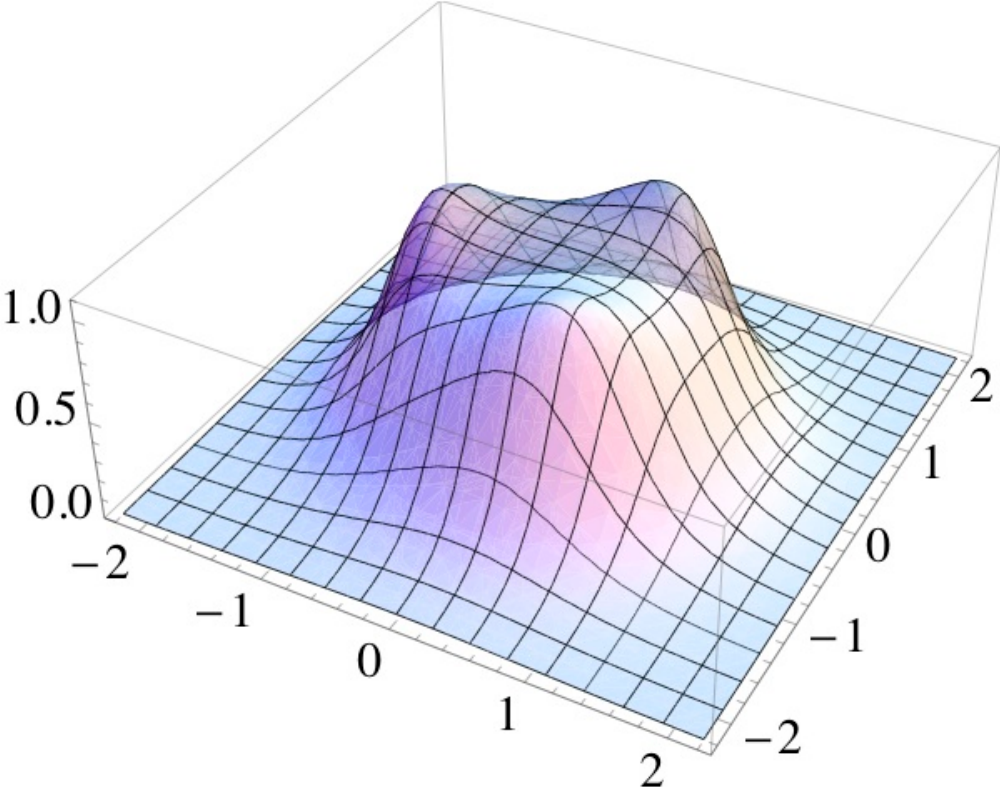}\includegraphics[width=0.3\textwidth]{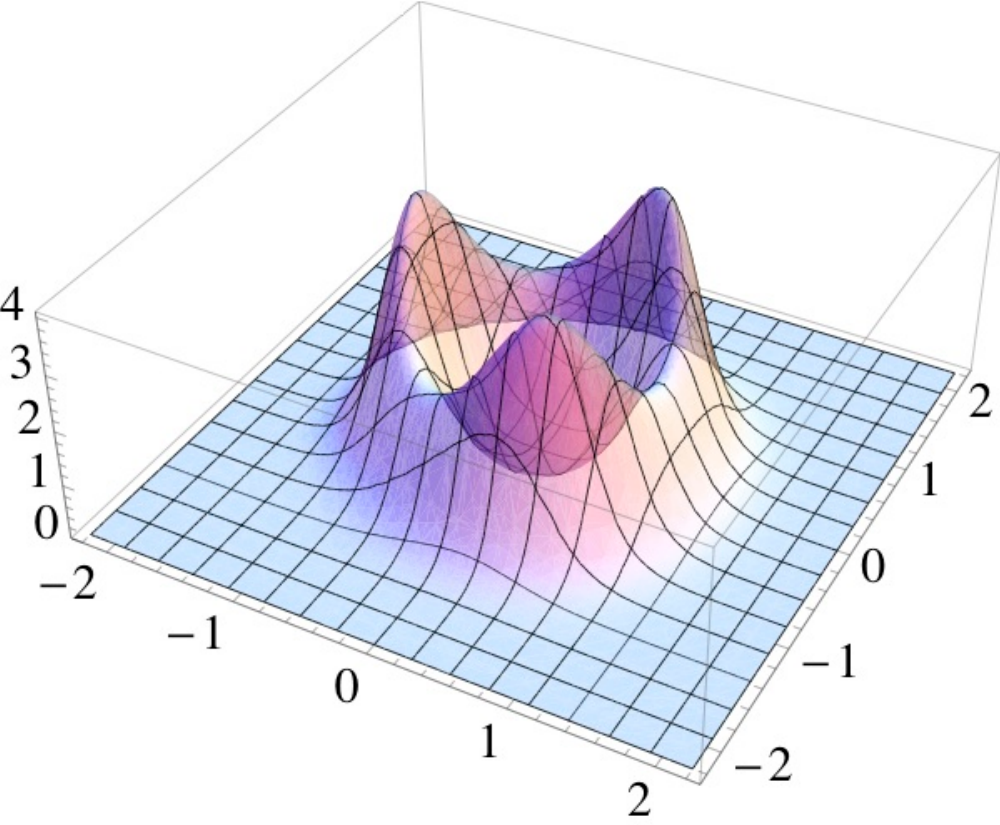}
\includegraphics[width=0.3\textwidth]{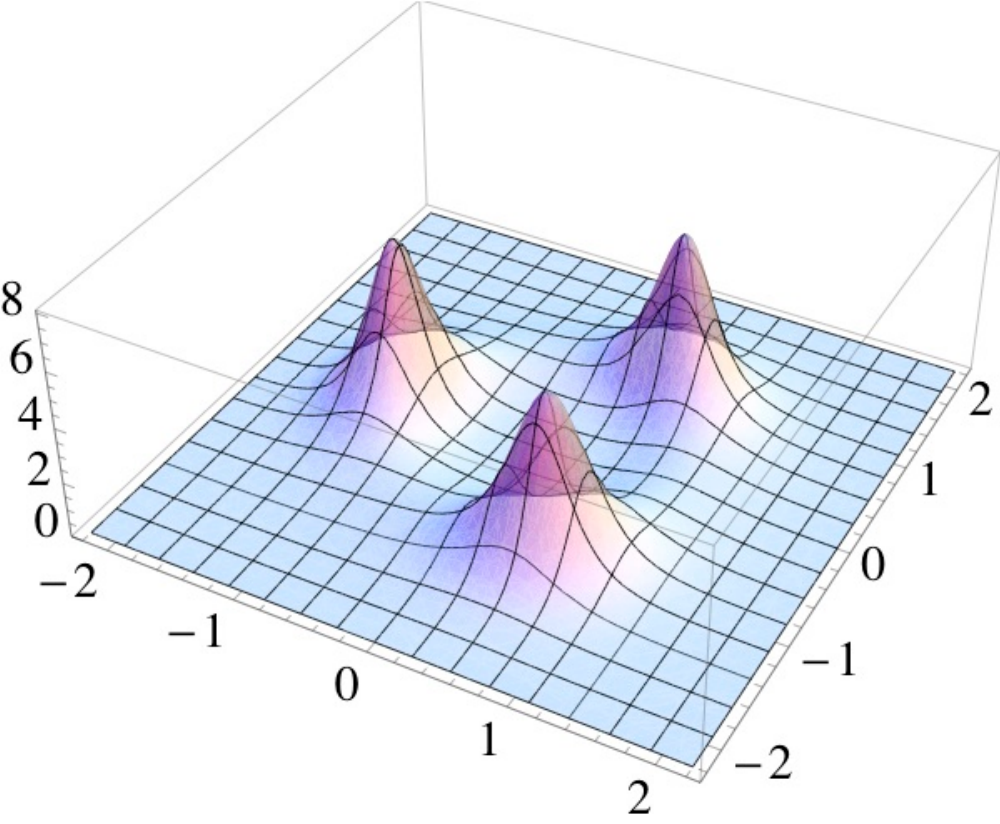}\includegraphics[width=0.3\textwidth]{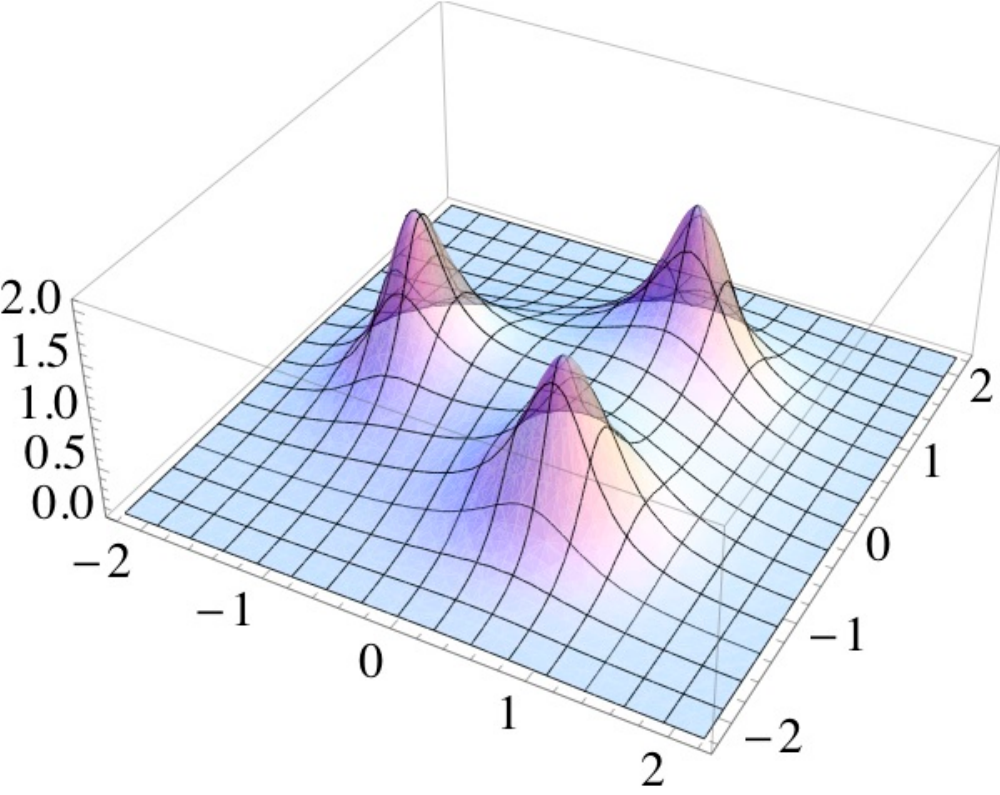}\includegraphics[width=0.3\textwidth]{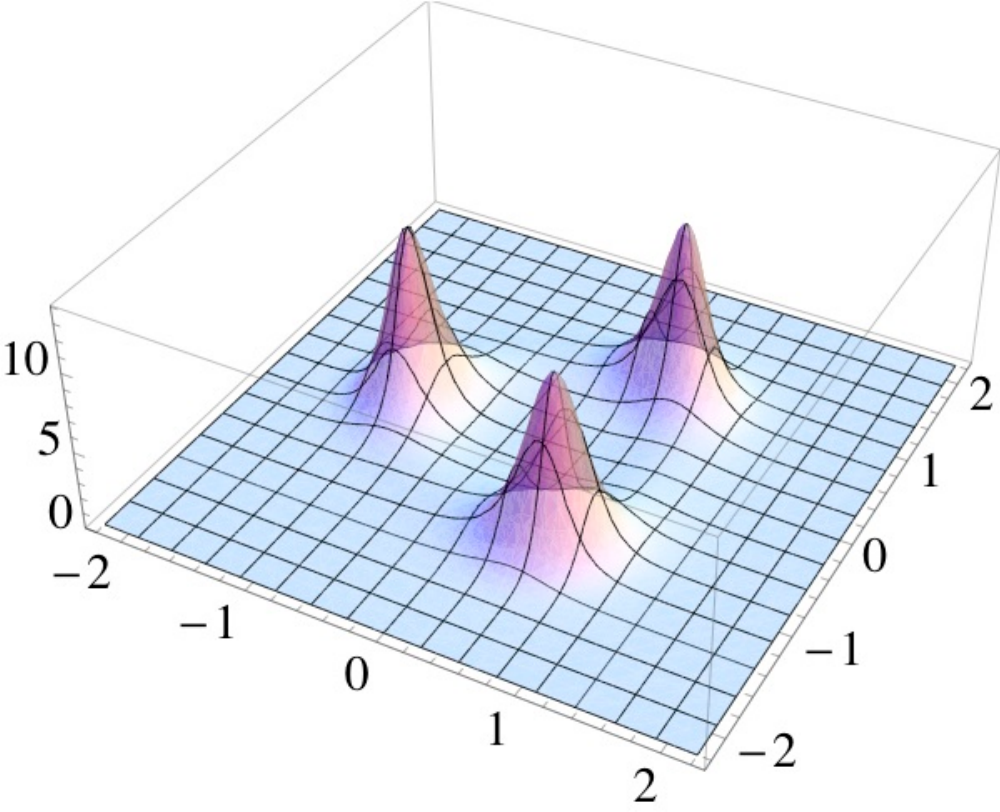}
\includegraphics[width=0.3\textwidth]{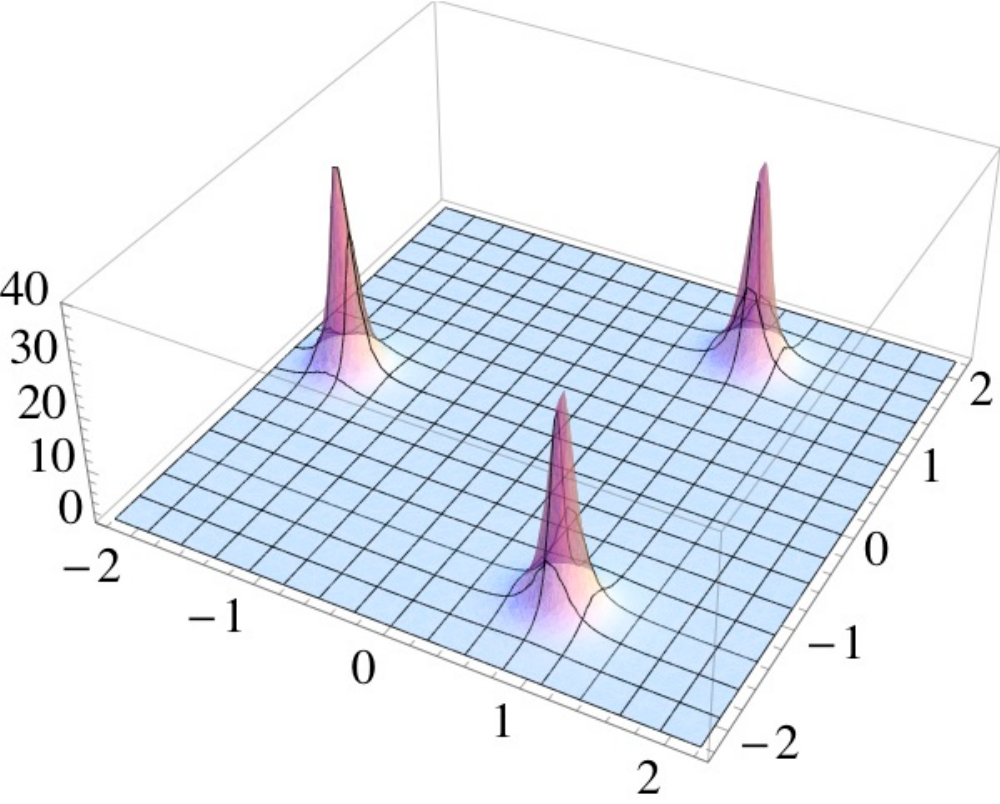}\includegraphics[width=0.3\textwidth]{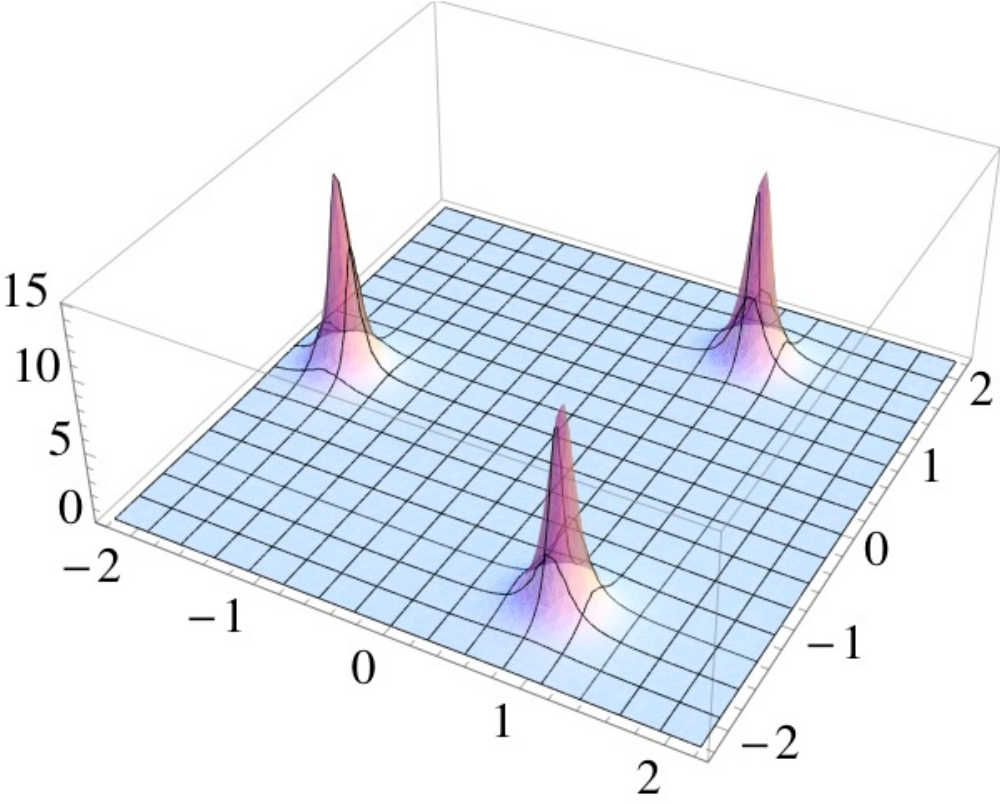}\includegraphics[width=0.3\textwidth]{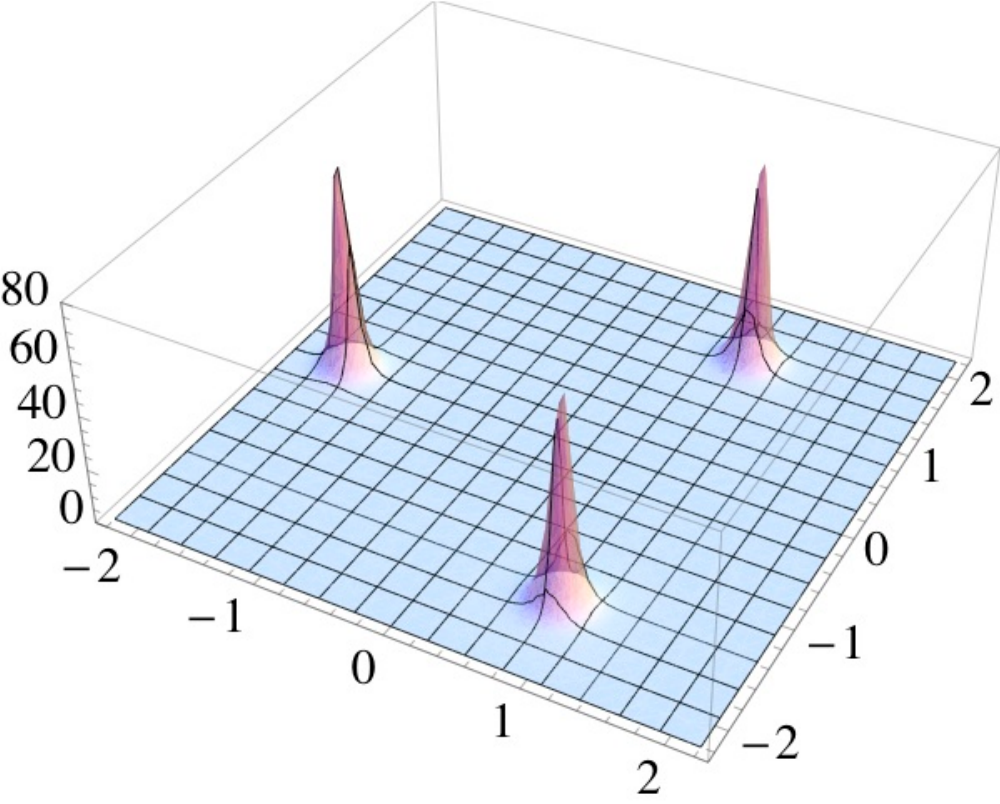}
\caption{The $Q=3$ solution with $a_2=0$. The functions $\mathcal{H}^{(1)}$ (first column), $\mathcal{H}^{(2)}$ (second column), $\mathcal{H}^{(3)}$ (third column) for $a_1=0$ (first row), $a_1=0.2$ (second row), $a_1=1$ (third row), $a_1=5.0$ (fourth row) and  $a_3=1.0$, $a_4=1.0$,    $k_1=1.0$, $k_2=2.0$, $y_+=0$.}
\label{threespiral}
\end{center}
\end{figure}

From these formulas it is  clear that $a_1$ is a crucial parameter which determines the distance between individual maxima. The leading behaviour for $|a_1|\gg1$ is $r^{(1)}=|a_1|^{1/Q}+\ldots$ and $r^{(3)}=|a_1|^{1/Q}+\ldots$ whereas $r^{(2)}=c|a_1|^{1/Q}+\ldots$ where $c$ depends on all other parameters $k_1$, $k_2$, $a_3$, $a_4$ and $Q$. An example of the energy density contributions is sketched in Fig.\ref{threespiral}. For $a_1=0$ the energy density contributions form a symmetric crater or a peak. When $a_1$ increases the picture distorts itself into three gradually emerging maxima which for $a_1\gg 1$ became three well localized peaks. These peaks rotate with the speed of light. When visualized in 3-dim. space the positions of maxima of the energy density take the form of three rotating spirals. The calculation of the energy per unit length gives
\br
\frac{1}{\pi}\int_{R^2}dx^1dx^2\mathcal{H}^{(1)}=Q,\quad \frac{1}{\pi}\int_{R^2}dx^1dx^2\mathcal{H}^{(3)}=Q\frac{k_2^2a_3^2a_4^2}{(a_3^2+a_4^2)^2}=k_2\mathcal{Q}^{(2)},\label{integrals}
\er  
where $Q_{\rm Top}=Q$ and $\mathcal{Q}^{(2)}$ is the Noether charge. The last formula can be obtained directly from (\ref{relation}) for $\lambda=k_2$ and $a=a_3/a_4$. 

\begin{figure}
\begin{center}
\includegraphics[width=0.55\textwidth]{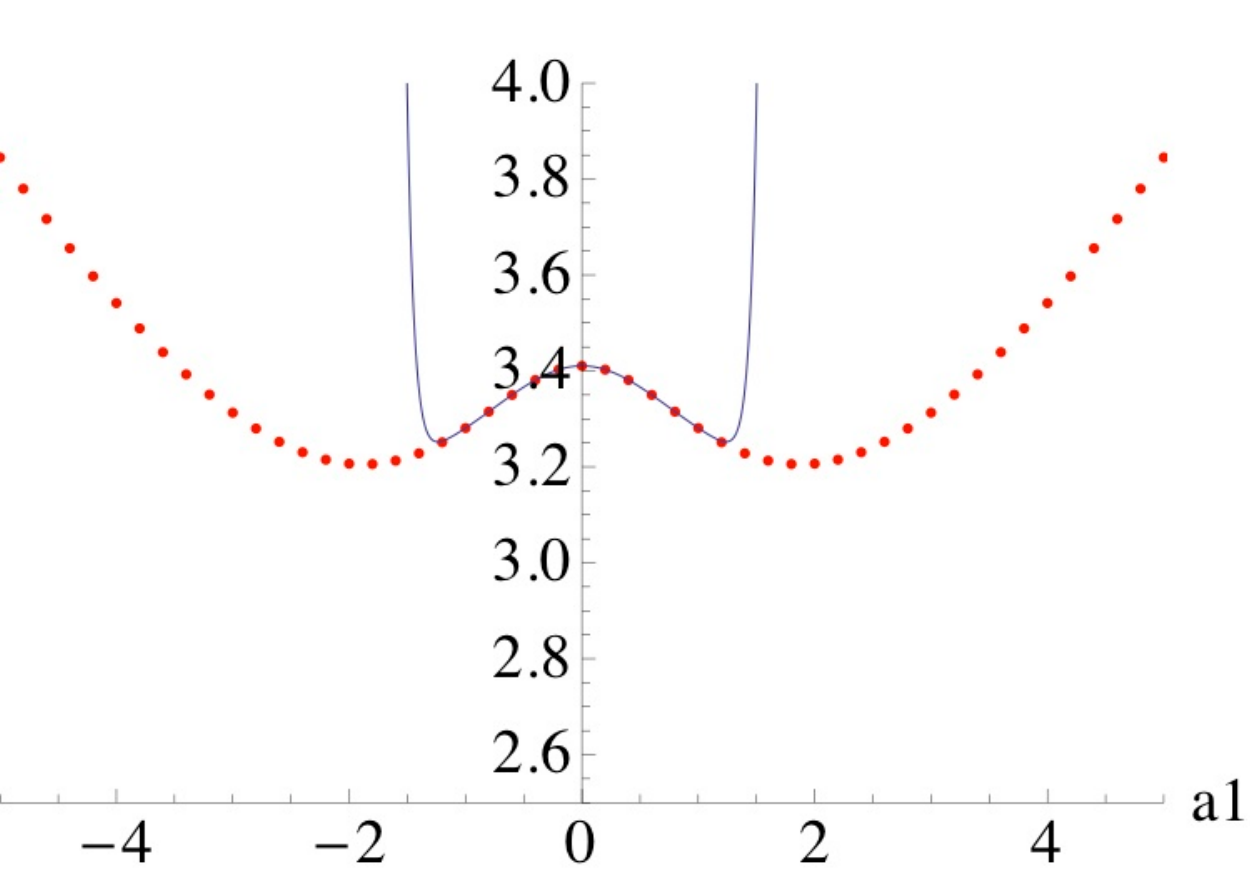}\caption{The case $a_2=0$. The integral $\int dx^1dx^2\mathcal{H}^{(2)}$ as the function of $a_1$ for $a_3=1$, $a_4=0.8$  $k_1=1.0$, $k_2=2.0$. The numerical integration are given by points whereas the curve shows a few first terms of expansion at $a_1=0$. The expansion converges for $|a_1|<1$.}
\label{I2_1}
\end{center}
\end{figure}

The quartic contribution vanishes in three cases: $a_3=0$, $a_4=0$ and $k_2=0$. For each of those cases there is effectively only one $u_i$ that matters, which  means that the target space reduces from $CP^2$ to $CP^1$. It is important to stress that both integrals (\ref{integrals}) do not depend on $a_1$ and therefore also on the distance between vortices. The only dependence comes from the integral  $\int dx^1dx^2\mathcal{H}^{(2)}$. Unfortunately we cannot calculate this integral analytically. The numerical integration, see Fig.\ref{I2_1}, shows that the energy per unit length takes a minimal value for some non-zero $a_1^{min}$. When $a_3\rightarrow 0$ the position of the minimum also tends to zero, $a_1^{min}\rightarrow 0$. 

The approximate dependence on  $a_1$ can be calculated analytically  around $a_1=0$ by integrating coefficients of its Taylor expansion 
$$
\int dx^1dx^2\mathcal{H}^{(2)}=\sum_{n=0}^{\infty}\mathcal{I}_{n}a_1^{2n}.
$$
Note that only even coefficients $\mathcal{I}_{n}$ are non-zero. The results of integration give the following general expression 
\begin{eqnarray}
\mathcal{I}_{n}&\equiv&\frac{(-1)^{n+1}\left(-\frac{1}{Q}\right)^2_{n+1}}{\left(n-\frac{1}{Q}\right)(n!)^2}\frac{\Gamma\left(1+\frac{1}{Q}\right)\Gamma\left(1-\frac{1}{Q}\right)}{(a_3^2+a_4^2)^{n+2-\frac{1}{Q}}}\times\nonumber\\
&\times&\left[k_1^2n^2Q^2(a_3^2+a_4^2)^2+k_2^2a_3^2(nQ-1)[(nQ-1)a_3^2-Qa_4^2]-\right.\nonumber\\
&-&\left.2k_1k_2a_3^2nQ(nQ-1)(a_3^2+a_4^2)\right],
\end{eqnarray}
where
$$
(a)_{n+1}\equiv a(1+a)(2+a)\ldots(n+a).
$$
is the Pochhammer symbol. The expansion converges for $|a_1|<1$. Unfortunately the analytical curve does not reach the minimum; nevertheless in the region of convergence it serves as a test confirming our numerical computations.  

\subsubsection{The case $N=Q$}

In the case $N=Q$ the energy density of all contributions $\mathcal{H}^{(a)}$, $a=1,2,3$, has an axial symmetry. In this case one has to exclude the values $a_1=\pm 1$  since they lead to the vanishing of $f_1$ for values of  $y_+$ such that $e^{ik_1y_+}=\pm1$. 
The integrals corresponding to $a=1$ and $a=3$ remain unchanged and they are given by formulas (\ref{integrals}).  Note that the corresponding energy densities (\ref{hamiltonians}) depend on $y_+$ via the periodic function $\cos{(k_1y_+)}$. 

It is important to stress that the formulas (\ref{integrals}) are the same in both cases $N=0$ and $N=Q$ but the energy density functions and their dynamics are completely different. In the first case there are rotating peaks whereas in the second one there are oscillating symmetric rings. The only dependence of the energy density on $a_1$ comes from the integral $\int dx^1dx^2\mathcal{H}^{(2)}$ which also depends on $y_+$. The integral can be computed analytically giving
\begin{eqnarray}
\frac{1}{\pi}\int_{R^2} dx^1dx^2\mathcal{H}^{(2)}=\Gamma_Q\frac{\alpha_Q}{\beta_Q},\label{intful}
\end{eqnarray}
where
\begin{eqnarray}
\Gamma_Q\equiv \frac{\pi}{Q^3}\sum_{n=1}^{Q-1}(-1)^n\left[\frac{1-(-1)^Q}{2}n-\frac{1-(-1)^n}{2}Q\right]\sin\left(\frac{n\pi}{Q}\right)\nonumber
\end{eqnarray}
and
\begin{eqnarray}
\alpha_Q&\equiv&a_3^2(a_3^2+Qa_4^2)k_2^2+a_1^2a_4^2k_1^2+a_1^2a_3^2(k_1-k_2)^2\nonumber\\
&+&a_1^2a_3^2a_4^2(2k_1^2-2k_1k_2+Qk_2^2)+\nonumber\\
&-&2a_1a_3^2k_2\left[a_4^2(k_1-Qk_2)+a_3^2(k_1-k_2)\right]\cos{(k_1y_+)}\nonumber\\
\beta_Q&\equiv&(a_3^2+a_4^2)^{2-\frac{1}{Q}}\left[1+a_1^2+2a_1\cos{(k_1y_+)}\right]^{1+\frac{1}{Q}}.\nonumber
\end{eqnarray}

\begin{figure}
\begin{center}
\includegraphics[width=0.5\textwidth]{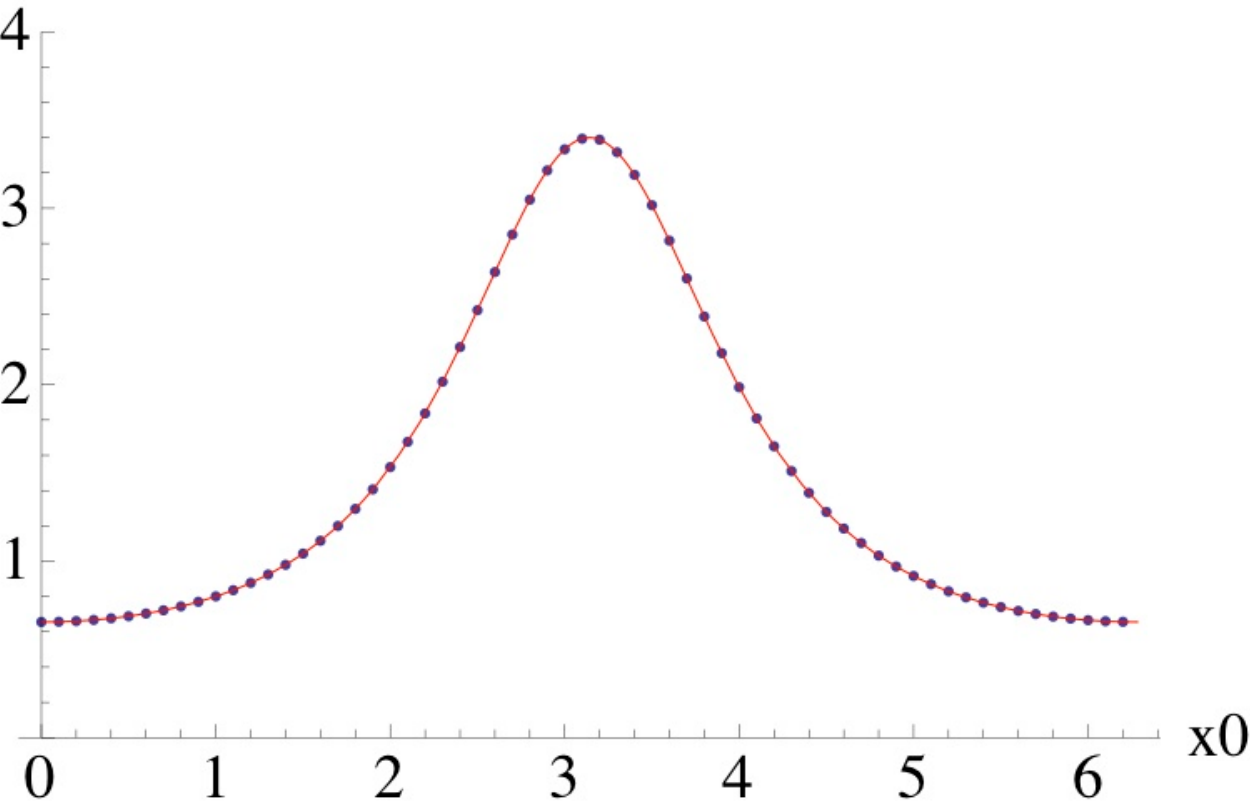}\includegraphics[width=0.5\textwidth]{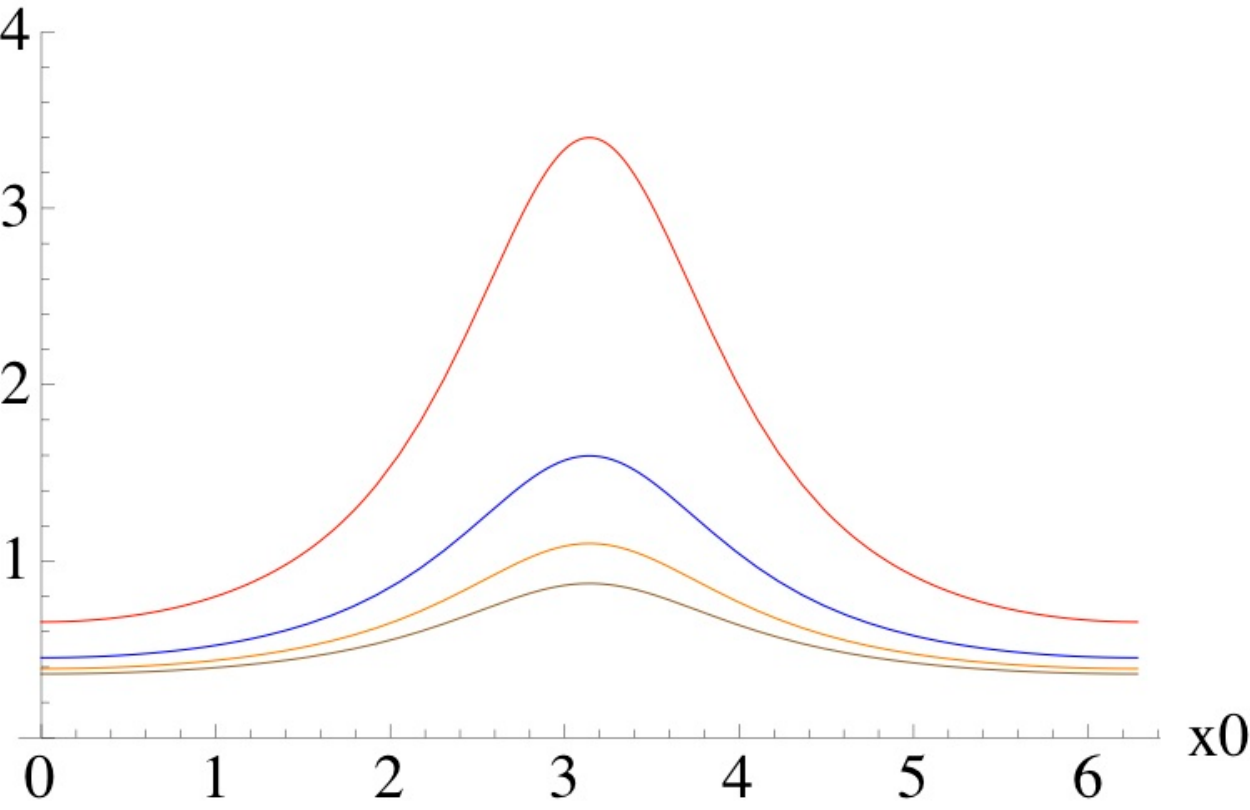}\caption{The integral $\frac{1}{\pi}\int_{R^2} dx^1dx^2\mathcal{H}^{(2)}$ as the function of time $x^0$. The left picture contains some numerical values (the points) and the analytical curve. It corresponds to the choice $Q=2$, $a_1=3$, $a_3=1$, $a_4=3.5$, $k_1=1$ and $k_2=2$. The right picture was obtained for the same values of the constants and various $Q$; from the top: $Q=2$, $Q=3$, $Q=4$, $Q=5$.}
\label{I2_oscilate}
\end{center}
\end{figure}

The integral (\ref{intful}) is in excellent concordance with its numerical estimate. Some corresponding curves are presented in Fig.\ref{I2_oscilate}. This integral is infinite for $a_1=1$ and  $k_1y_+=(2n+1)\pi$ or $a_1=-1$ and $k_1y_+=2n\pi$ where $n=0,\pm 1,\pm 2,\ldots$ Since we have excluded $a_1=\pm 1$ from the space of parameters the integral is always finite but it could take very high values close to the points where $f_1$ vanishes. The energy per unit lengths always tends to zero for $a_1\rightarrow \pm\infty$ where the leading behaviour is $\sim |a_1|^{-2/Q}$. This example shows that for some energy density contributions, like a quartic term which we discussed here, the integral over the $x^1x^2$ plane can be time independent in spite of the fact that the energy density is not. It is important to stress that there is no rotation here which explains the time independence of the energy per unit length. 

\subsection{Example 2}

Another interesting solution can be obtained by a straightforward generalization of the vortex solution presented in  \cite{CPNvortex}. Let us consider, for instance, the one parameter solution of the form
\br
u_j(z,y_+)=(z-\delta)^{n_j}(z+\delta)^{m_j}e^{ik_jy_+},\qquad j=1,2\label{multivortex}
\er 
where $\delta$ is some real number. For the values of $\delta$ being sufficiently  large  the energy density is localized around the points $z=\pm\delta$ taking the form of either peaks or craters.  As a vortex is not a point-like object in order to study the dependence of the energy per unit length on the separation between vortices one has to define what the expression ``the distance between vortices'' means.
In order to do so one needs to chose some characteristic parameters of the solution. In the case of the solution (\ref{multivortex}) we only have the points $z=\pm\delta$ in the complex plane. For $\delta\gg1$ the value $2\delta$ is a distance between centers of two local peaks (or craters) in the plot of the energy density.  However, for smalles $\delta$'s the picture of the energy density becomes more complicated and the reference to the points $z=\pm\delta$ as to the centers of the vortices is no longer valid. Thus, we can think of $2\delta$ as the distance between vortices only for separated vortices.

\subsubsection{Topological sector}

\begin{figure}
\begin{center}
\includegraphics[width=0.55\textwidth]{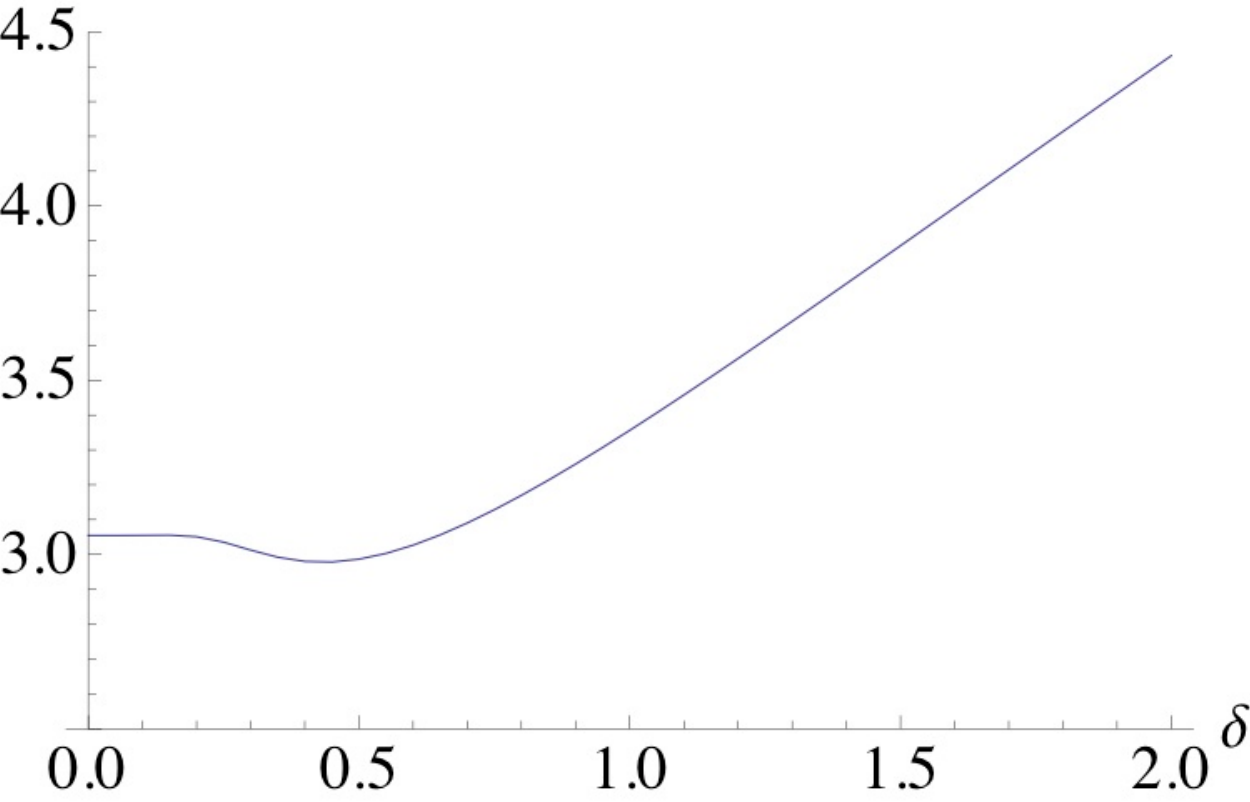}\caption{The integral $\frac{1}{\pi}\int_{R^2} dx^1dx^2\mathcal{H}^{(2)}$ as the function of $\delta$ for $n_1=n_2=-3$, $m_1=m_2=1$, $k_1=1$ and $k_2=2$. This is the only integral which depends on $\delta$ since the quartic contribution is topological for this case.}
\label{ex2_a}
\end{center}
\end{figure}

According to discussion in the previous section, for the solutions of the form $u_i(z,y_+)=v(z)e^{ik_iy_+}$ the quartic term contribution to the energy density has a topological nature. Taking $n_1=n_2\equiv n$ and $m_1=m_2\equiv m$ one can replace the solution by the equivalent one
$$
\tilde u_1=(z-\delta)^{-n}(z+\delta)^{-m}e^{-ik_1y_+},\qquad \tilde u_2=e^{i(k_2-k_1)y_+}
$$
which gives
$$
\int_{R^2}dx^1dx^2\mathcal{H}^{(3)}=Q_{\rm Top}\frac{(k_1-k_2)^2}{4},
$$
where for $\delta > 0$, $Q_{\rm Top}=|n+m|$ for $nm>0$ and $Q_{\rm Top}={\rm max}(|n|,|m|)$ for $nm<0$. When $m=0$ then  $Q_{\rm Top}=|n|$ and vice versa. The integral does not depend on $\delta$ which is a counterpart of the parameter $a_1$ from Example 1. Fig.\ref{ex2_a} shows the integral $\frac{1}{\pi}\int_{R^2} dx^1dx^2\mathcal{H}^{(2)}$ which is the only contribution to the total energy per unit length which depends on the distance between vortices. For solutions of the form (\ref{multivortex}) not all combinations $n_i$ and $m_i$ lead to the finite energy per unit length. In fact a grid of points corresponding to acceptable combinations of exponents forms a quite complex structure but we shall not discuss this problem here because it is beyond the scope of present paper.

\subsubsection{Outside the topological sector}

\begin{figure}
\begin{center}
\includegraphics[width=0.5\textwidth]{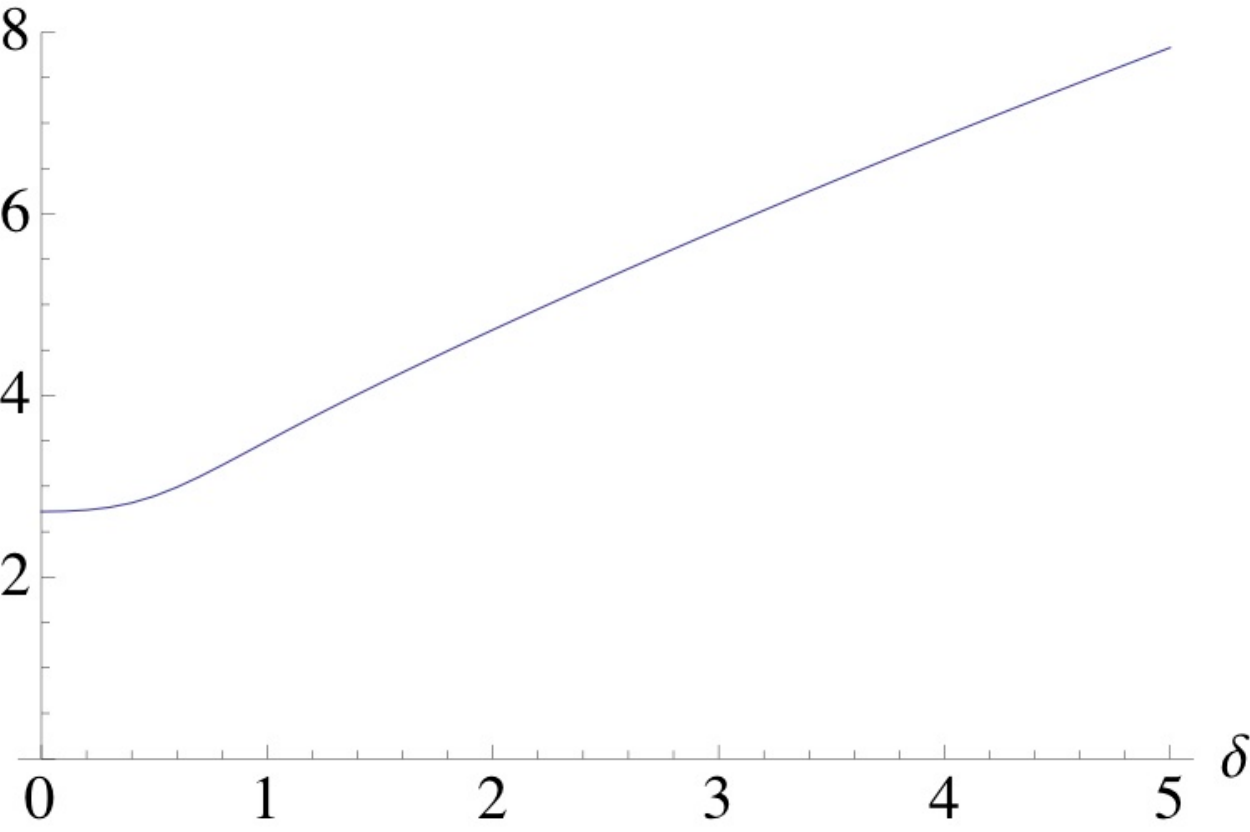}\includegraphics[width=0.5\textwidth]{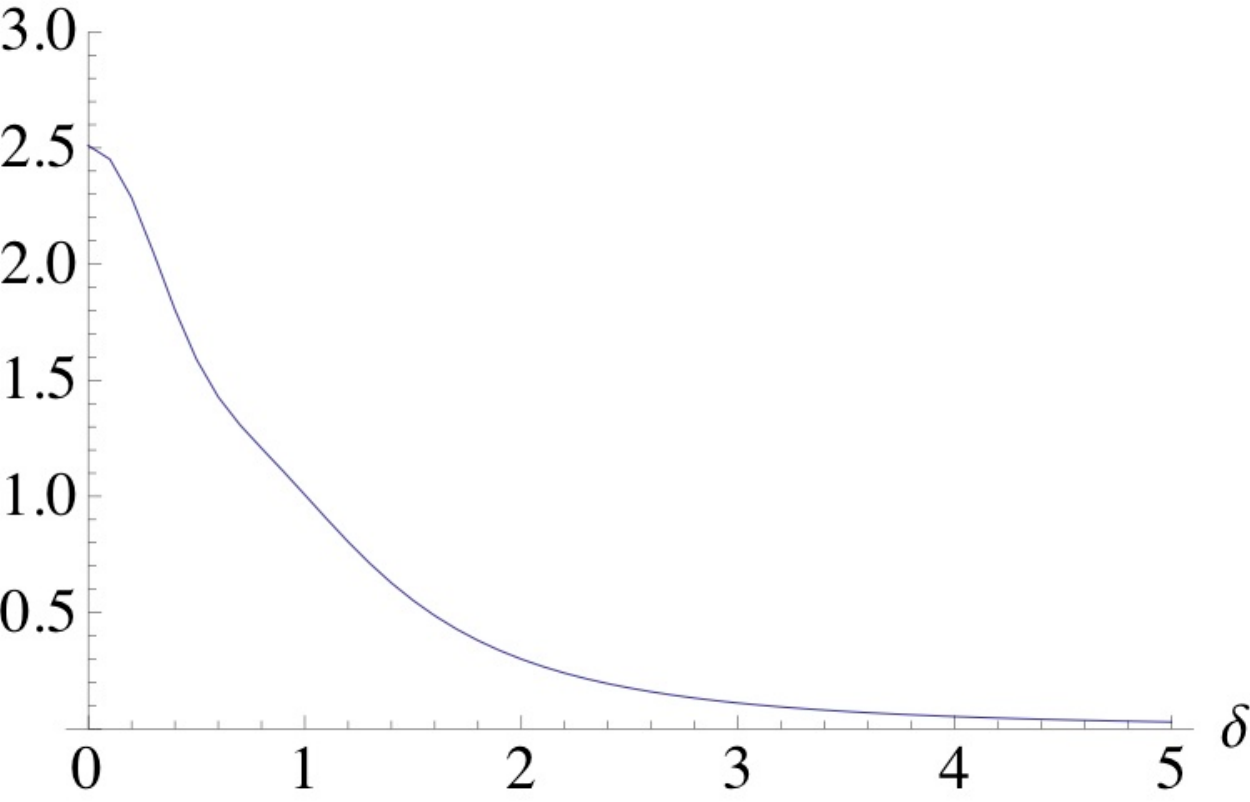}\caption{The integral $\frac{1}{\pi}\int_{R^2} dx^1dx^2\mathcal{H}^{(2)}$ (left) and $\frac{1}{\pi}\int_{R^2} dx^1dx^2\mathcal{H}^{(3)}$ (right) as the function of $\delta$ for $n_1=n_2=-1$, $m_1=-1$, $m_2=3$, $k_1=1$ and $k_2=2$. The quadratic contribution to the energy density has minimum at $\delta=0$.}
\label{ex2_b}
\end{center}
\end{figure}

\begin{figure}
\begin{center}
\includegraphics[width=0.5\textwidth]{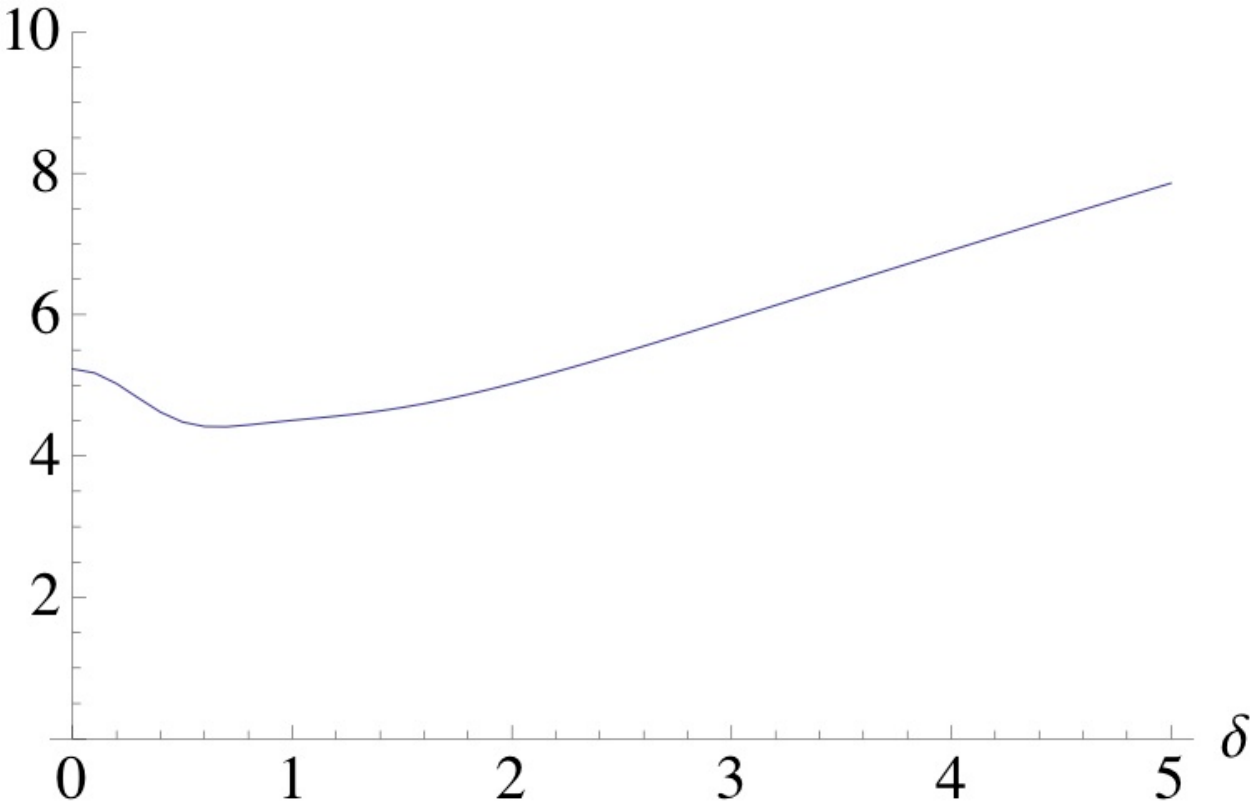}\includegraphics[width=0.5\textwidth]{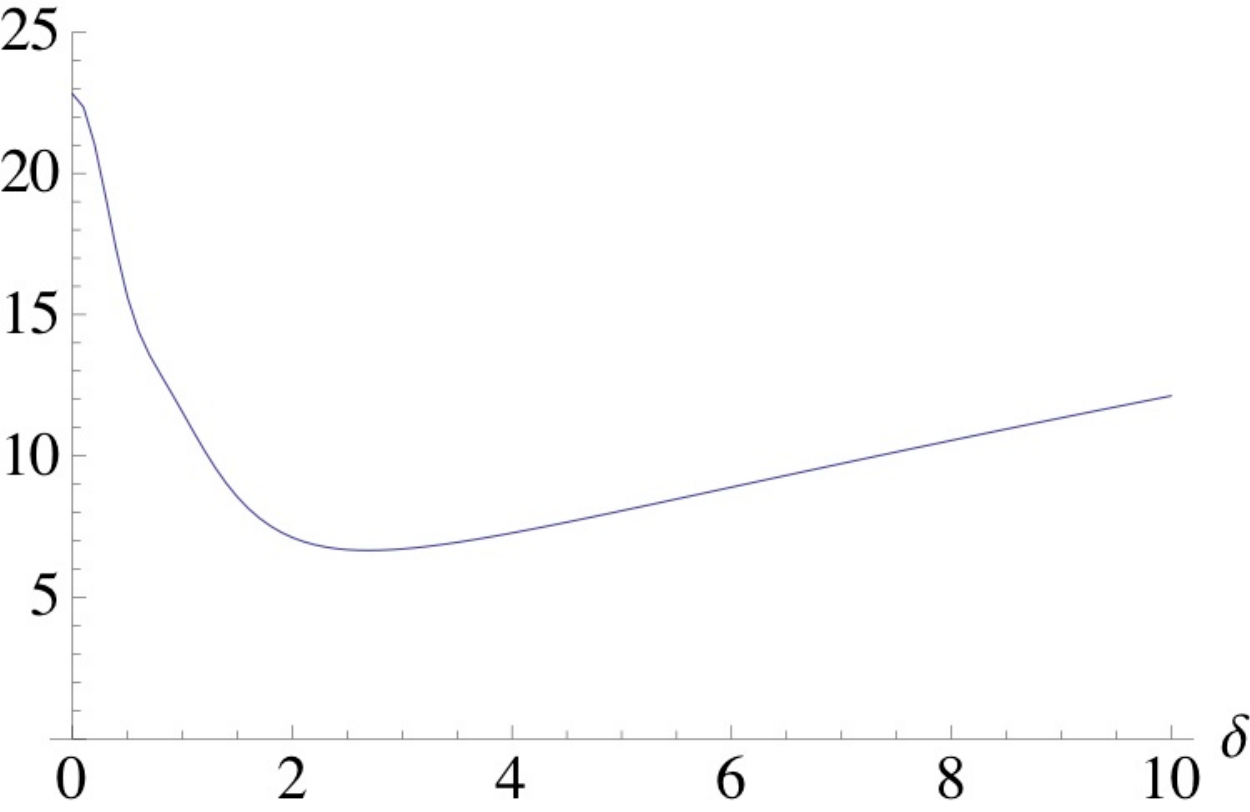}\caption{The integral $\frac{1}{\pi}\int_{R^2} dx^1dx^2(\mathcal{H}^{(2)}+\mathcal{H}^{(3)})$ (left) and $\frac{1}{\pi}\int_{R^2} dx^1dx^2(\mathcal{H}^{(2)}+8\mathcal{H}^{(3)})$ (right) as the function of $\delta$ for $n_1=n_2=-1$, $m_1=-1$, $m_2=3$, $k_1=1$ and $k_2=2$. The minimum $\delta_{min}>0$ appears as the effect of mutual interaction of the quadratic and the quartic term.}
\label{ex2_c}
\end{center}
\end{figure}

The common property of all solution having the quartic term topological is that the minimum of the total energy density is determined only by properties of the integral $\frac{1}{\pi}\int_{R^2} dx^1dx^2\mathcal{H}^{(2)}$. Because of this the position of the minimum does not depend on the coupling constants. When the quartic term cannot be reduced to the topological case the minimum (if it exists) depends on the mutual relation between quadratic and quartic terms. The position of such a minimum is a function of the ratio $8(\gamma-\beta)/M^2$. This possibility is new since the quartic term for the model with a target space being $CP^1$ is identically zero if one imposes the integrability conditions (\ref{constraint}). As an example we consider a solution of the modified Skyrme- Fadeev model in a $CP^2$ target space. Such a solution corresponds to (\ref{multivortex}) for the choice $n_1=n_2=-1$, $m_1=-1$, $m_2=3$, $k_1=1$ and $k_2=2$. Fig.\ref{ex2_b} shows two contributions to the energy density. The quadratic contribution leeds to energy per unit length that increases with a distance between two multi-vortices. This contribution has a minimum at $\delta=0$. The contribution to the energy per unit length given by the quartic term has maximum for $\delta=0$ and it decreases as $\delta$ tends to infinity what means that the quartic term leads to a  repulsive interaction between two multi-vortices. Fig.\ref{ex2_c} shows the
 sum of quadratic $\mathcal{H}^{(2)}$ and $\mathcal{H}^{(3)}$ contributions to the total energy per unit length
$$
\frac{1}{\pi}\int_{R^2} dx^1dx^2\left(\mathcal{H}^{(2)}+\frac{8(\gamma-\beta)}{M^2}\mathcal{H}^{(3)}\right).
$$
In both cases the combination of the terms leads to a minimum situated at some $\delta>0$. The left picture corresponds to the choice $(\gamma-\beta)/M^2=1/8$ whereas the right picture shows the combination with $(\gamma-\beta)/M^2=1$.

\section{Concluding remarks}

In this paper we have shown that holomorphic solutions of the $CP^N$ model are also 
solutions of the extended $CP^N$ Skyrme-Faddeev model. In fact, these are very special solutions since they belong to an integrable sub-model of  the $CP^N$ Skyrme-Faddeev theory, defined by the constraints (\ref{constraint}), and the condition on the coupling constants given by $\beta\,e^2+\gamma\,e^2=2$. The constraints (\ref{constraint}) imply that such sub-model possesses   an infinite number of conserved currents. As the total energy of these solutions is infinite these solutions should be interpreted as describing time dependent vortices, and they generalize the results obtained in \cite{CPNvortex,fkz1,fkz2}. 

In this paper we have studied the properties, and in particular the dynamics of these multi-vortex solutions of the $CP^N$ Skyrme-Faddeev model. The structure of the solutions is very complex and diverse, and we have in fact considered only some types of solutions which are  inte\-res\-ting from the physical point of view. We have shown for instance that in some cases  one of the terms in the energy density is related to the Noether charge of the model, and in some other cases the energy density can be factorized in the product of two terms where one of them is the topological charge. 

It has been put forward in \cite{Kondo} that the $CP^N$ Skyrme-Faddeev model might describe some aspects of the low energy (strong coupling) regime of the pure $SU(2)$ Yang-Mills theory. If that is indeed the case, the solutions constructed in this paper certainly must play a role, describing some type of low energy excitations of the Yang-Mills theory. For those reasons more research work is needed to get a better
understanding of these interesting phenomena.

\vspace{1cm}

{\bf Acknowlegment:} L.A. Ferreira and W.J. Zakrzewski would like to thank
the Royal Society (UK)
for a grant that helped them in carrying out this work. L.A. Ferreira
is partially supported by CNPq (Brazil) and P. Klimas is supported by
FAPESP (Brazil).

\end{document}